\documentclass[aps,prd,preprint,groupedaddress,showkeys,showpacs]{revtex4-1}
\usepackage{graphicx}
\usepackage{dcolumn}
\usepackage{diffcoeff} 
\usepackage[italicdiff]{physics}
\usepackage{bm}%
\usepackage[dvipsnames]{xcolor}
\usepackage{epstopdf}
\usepackage[toc]{appendix}
\usepackage{float}
\usepackage{amsmath}
\usepackage[caption = false]{subfig}
\begin{document}

\preprint{APS/123-QED}

\title[Study of GR shock propagation in NSs]{The importance of general relativistic shock calculation in the light of neutron star physics}

\author{Anshuman Verma}
 \email{anshuman18@iiserb.ac.in}
 \affiliation{Department of Physics, Indian Institute of Science Education and Research Bhopal.
 	}

\author{Ritam Mallick}
 \email{mallick@iiserb.ac.in}
\affiliation{Department of Physics, Indian Institute of Science Education and Research Bhopal.
	}%

\date{\today}

\begin{abstract} 
Numerical simulation of hydrodynamic equations forms the central part of solving various modern astrophysical problems. In the case of shocks, one can have either dynamical equations or jump conditions (the conservation equations without any time evolution). The solution of the jump condition in curve space-time is derived and analyzed in detail in the present work. We also derive the Taub adiabat or combustion adiabat equation from the jump condition. We have analyzed both time-like and space-like shocks in the present work. We find that the change in entropy for the weak shocks for curved space-time is small similar to that for flat space-time. We also find that for general relativistic space-like shocks, the Chapman-Jouguet point does not necessarily correspond to the sonic point for downstream matter, unlike the relativistic case. To analyze the shock wave solution for the curved space-time, one needs the information of metric potentials describing the space-time, which for the present work is taken to be a neutron star. We assume that a shock wave is generated at the centre of the star and is propagating outward. As the shock wave is propagating outwards, it combusts nuclear matter to quark matter, and we have a combustion scenario. We find that the general relativistic treatment of shock conditions is necessary to study shocks in neutron stars so that the results are consistent with the solution of the TOV equation while calculating the maximum mass for a given equation of state. We also find that with such general relativistic treatment, the combustion process in neutron stars is always a detonation.

\end{abstract}

\maketitle

\section{Introduction}

Shock waves have seen numerous applications in the field of fluid, astrophysics, plasma physics, and heavy ion collision, to name a few. Shock waves are a common phenomenon in astrophysics and are generated in high-energy processes like a supernova, gamma-ray bursts, binary mergers, etc. \cite{Bird,Abe,Weaver,Giannios,sri}. They are responsible for particle acceleration in high energetic processes like gamma-ray bursts, fast radio bursts, and solar activity \cite{bland,kirk,abr,dejan}. It is also postulated that the cosmic rays are particles accelerated by shocks after a supernova \cite{bell1,bell2,Mor,Lam}. Shock dissipation mechanisms can be radiative in nature under certain conditions and are termed radiation-mediated shocks and are associated with a supernova, gamma-ray bursts, and neutron star mergers \cite{Budnik,Lundman1,Lundman2,Naker}. There is another prospect of collision-less shocks, which are mediated by collective plasma effects like earth bow shock in solar winds \cite{bisk,treu,marc}. There have also been works using shock waves in a heavy-ion collision where a phase transition (a detonation) from quark-gluon plasma to hadronic matter is accompanied by a shock \cite{csernai,suho,khle,shur}. A similar but opposite shock-induced phase transition from hadronic matter to quark matter at the core of neutron stars has also gained attention in recent years \cite{collin,marr,drago1,ghosh,drago2,m1,m2,prasad1,prasad2}.

The earliest concept of shocks dates back to Euler's equations\cite{demet,tai}. Shock waves or sharp shock discontinuity are generated when the fluid velocities exceed the local sound velocity. Around the discontinuity surface, the thermodynamic variables vary discontinuously; however, they are related by the conservation of mass, energy, and momentum flux across the surface \cite{book1,liva}. The special relativistic (SR) theory of shock waves was first introduced by Taub \cite{taub} and was later improved by Landau and Lifschitz \cite{book2}. A more elaborate and detailed calculation of shock waves was later done by Lichnerowich \cite{lich}, and Thorne \cite{thorne}. Thereafter, the concept of SR shock waves was abundantly used in astrophysics to study various phenomena. Basically, there are two ways to employ the concept of shock in a problem: the first way is using the dynamic Euler's equations to study a phenomenon and have a complete evolutionary picture of the problem \cite{trac,dani,prasad1}. The other is to employ the so-called jump condition (also derived from Euler's equations) to analyze a problem \cite{kent,gil,abd}. The latter is a simpler method where we lose the dynamic evolution of the shock but have a really good estimate of how the thermodynamic variables vary across the shock, which can further be used to deduce several properties of the system.

In recent years general relativistic (GR) hydrodynamic equations are numerically solved in the problems like a supernova, binary neutron star mergers (BNSM) \cite{drake,ott,smith,faber,shib}. As they are highly nonlinear and coupled equations, to solve the system in such astrophysical scenarios, we need to solve Einstein equations and hydrodynamic equations simultaneously. They can only be solved numerically, and for that purpose, we need to discretize both space and time, and the concept of numerical relativity develops. In connection to binary neutron star merger, Most et al. \cite{most} recently simulated such a system to check for PT in post-merger product of BNSM and obtained its gravitational wave (GW) signature. Fischer et al. \cite{fischer} studied the explosions of massive stars, which are triggered via quark-hadron PT during the early post-bounce phase of core-collapse SN. There are also works where one solves hydrodynamic equations involving weak interaction in isolated neutron stars \cite{neibergal,ouyed} to account for PT in them. Although the above works solve hydrodynamic equations in a complicated setup, they do not set a shock problem in them (Riemann problem). Shock evolution in connection with PT in neutron star in isolated neutron star has also been studied, but they are done mostly in a 1-dimensional setup \cite{prasad1,prasad2,ritam-slow}. 

Ideally, the dynamic equations are suited to study the complete evolution of the system but setting up the shock problem along with phase transition in GR hydrodynamics (GRHD) is very complicated. On the other hand, kinematic shock equations (the jump conditions or the conservation conditions across the shock) are easy to solve as they are algebraic equations connecting thermodynamic properties across the shock front \cite{taub}. Such analysis does not give the evolutionary picture of the system, but many important properties and features of the system can be extracted from them. However, in the literature, the jump conditions that are mostly used are SR in nature \cite{csernai1,thorne}. There has not been much work to study and analyze the GR jump conditions, which can also provide important insight into more complex problems. In this regard, there have been only a handful of work by smoller \cite{smoller}, MTW \cite{mtw}, and Font \cite{font}. However, their results are seldom utilized in the study of astrophysical shock waves. 

In this paper, we have derived the GR jump conditions and Taub adiabat equation. To add, shock wave discontinuity can be both space-like (SL) and time-like (TL) (as shown by Csernai \cite{book3}), and we have studied both types of shocks. The paper is arranged in the following way: Section II discusses the general theory of shock jump conditions and Taub adiabat. In section III, we particularly study the conditions for the weak shock wave, and in section IV, we analyze the Chapman-jouguet (CJ) point for strong shocks. Next, we use the formalism of GR shocks to study phase transition in neutron stars in section V. And finally, in section VI, we summarize our results and draw conclusions from them.

\section{General relativistic jump condition and Taub adiabat}
In the present analysis, we have considered a perfect fluid having no heat exchange and without any shear stresses and viscosity. The thermodynamic quantities of the fluid vary discontinuously at the shock front. The following symbols denote thermodynamic quantities measured in the local rest frame of the fluid,

\begin{align*}
    &\mu = \diffp{\epsilon}n[s] = \frac{(p+\epsilon)}{n} = (p+\epsilon)V \Rightarrow \text{ chemical potential}\\
    &v_s = \Bigg[\diffp{p}\epsilon[s]\Bigg]^{1/2} \Rightarrow \text{  3-velocity of sound relative to fluid } \\
    &\diffp{(\mu V)}p[s] = V^{2} - V^{2}\mu\diffp{n}p[s] = V^{2}[1-\frac{1}{v_s^2}]\\
    &u^{\mu}_s =\frac{(1,v_s,0,0) }{\sqrt{1 - v^{2}_s}} \Rightarrow \text{4-velocity of sound w.r.t. to fluid in SR} \\
    &u^{\mu}_{s} = \frac{(1,v_s,0,0)}{[e^{2\phi}-e^{2\Lambda}v_s^2]^{1/2}} \Rightarrow \text{4-velocity of sound w.r.t. fluid in GR} \\
\end{align*}

The first law of thermodynamics is given in the form
\begin{align*}
    \dd \mu = V \dd p + T \dd s
\end{align*}
and is valid in both curve and flat space-time (ST).


Considering f(r,t) to be a general variable (like density, pressure, specific energy, etc.) that satisfies the conservation equations (mass, momentum, and energy), the general form of the conservation equation can be written as
\begin{align}
    \diffp{f(r,t)}t + \nabla.\phi_f = S_f.
\end{align}
To be precise the above equation can be expressed as
\begin{align}
    \frac{\partial}{\partial t}\textit{ (Density of the Quantity)} + \nabla.\textit{(Flux of the Quantity)} \nonumber \\
    = (Source - Sink)
    \label{partial}
\end{align}


\begin{figure}
\centering
\hspace{0.5cm}
  \includegraphics[scale=0.8]{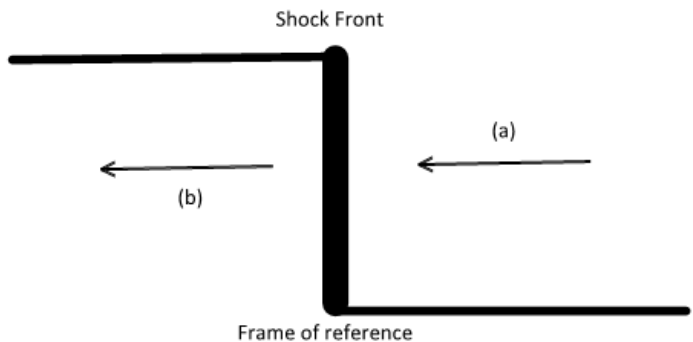}
  \caption{Schematic diagram of the shock front as a frame of reference where "a" denotes the upstream and "b" denotes the downstream matter w.r.t. the front.}
  \label{sides}
\end{figure}


To derive the jump conditions (conservation conditions across the shock) for a curved ST, one needs to define a metric that describes the curvature of the ST.
For GR, the line element of a spherically symmetric ST is given by 

\begin{equation}
    ds^2= g_{\mu\nu}dx^{\mu}dx^{\nu}=- e^{2\phi(r)}dt^2 + e^{2\Lambda(r)}dr^2 + r^{2}d\Omega^{2}
\end{equation}
where $d\Omega^{2} = d\theta^{2} + \sin^{2}({\theta}) d\Phi^{2}$ and $g_{\mu\nu}$ is a general spherically symmetric metric. Along with the metric, we also need to define the energy-momentum tensor, which contains information about the matter. Considering the form of the energy-momentum tensor as
 
\begin{equation*}
    T^{\mu\nu} = wu^{\mu}u^{\nu} + pg^{\mu\nu}  
\end{equation*}
where, $w$ (work function) = energy density ($\epsilon$ ) + pressure ( p) and the 4-velocity in curved ST is given by
\begin{equation}
    u^{\mu} = \dfrac{dx^{\mu}}{ d\tau}  = \dfrac{dx^{\mu}dt}{dt d\tau}\nonumber 
    \end{equation}    
where, the 4-position vector in spherical polar coordinate is $x^{\mu}=(t,r,\theta,\Phi)$ and the norm of $u^\mu$ is given by $g_{\mu\nu}u^\mu u^\nu = -1 $.

Considering the front to be moving along the radial direction, the 4-velocity of fluid particles is given by
\begin{equation}
    u^{\mu} = \gamma_g(1,v,0,0)
\end{equation}
where, $\theta $  and $\Phi $ are constant, $\gamma_g=\frac{1}{[e^{2\phi}-e^{2\Lambda}v^2]^{1/2}}$, and $v=\dfrac{dr}{dt}$ is radial velocity.


Since we are considering the curved ST, the partial derivative changes to the covariant derivative. Therefore, the energy-momentum and particle conservation equations in the curved ST can be generalized as

\begin{equation}
 \nabla_\nu T^{\mu\nu} = 0   
 \label{eng-mom}
\end{equation}
    
\begin{equation}
   \nabla_\mu (n u^{\mu}) = 0 . 
\end{equation}    
      
The first equation is the energy-momentum conservation equation, consisting of 4 equations, of which two are of our interest, and the other two are redundant. Therefore, using energy-momentum tensor expression for curved ST, the two equations of interest can be explicitly written. The first equation (of eqn \ref{eng-mom}) becomes

\begin{align}
    \nabla_{0}T^{00} + \nabla_{1}T^{01} = 0 \nonumber\\
    \Rightarrow \partial_{0}T^{00} + \partial_{1}T^{01} + T^{01}(3 \phi'(r) + \Lambda'(r)) =0 \nonumber \\
    \Rightarrow  \frac{\partial_{0}T^{00}}{T^{01}} + \partial_{1}\Big(\log(T^{01}) + 3 \phi(r) + \Lambda(r)\Big) = 0 \nonumber \\
    \Rightarrow \frac{\partial_{0}T^{00}}{T^{01}} + \partial_{1}\Big(\log(T^{01} e^{3 \phi(r) + \Lambda(r)})\Big) = 0  \label{1st}
\end{align}

Similarly, the second equation of interest (of eqn \ref{eng-mom}) is 
\begin{align}
    \nabla_{0}T^{10} + \nabla_{1}T^{11} = 0 \nonumber\\
    \Rightarrow \partial_{0}T^{10} + \partial_{1}T^{11} + 2 T^{11}\Lambda'(r) +  \nonumber \\
    T^{11}\phi'(r)\big(1 + \frac{T^{00} e^{2 (\phi(r) - \Lambda(r))}}{T^{11}}\big) =0\nonumber
\end{align}

The above equation can be further simplified as 

\begin{align}
     &\frac{\partial_{0}T^{10}}{T^{11}} + \partial_{1}\Big[\log(T^{11}) + 2  \Lambda(r) + \phi(r) \frac{T^{00}e^{2 (\phi(r) - \Lambda(r))}}{T^{00}} \nonumber \\
     &-  \int_{a}^{b}\phi(r)\Big(\frac{T^{00} e^{2 (\phi(r) - \Lambda(r))}}{T^{11}}\Big)'\,dr\ \Big] = 0. 
     \label{int}
\end{align}

Solving the integral of the above eqn \ref{int} (integration by parts), we have two terms coming from the integral. The first term coming from the integral cancels with the 4th term of eqn \ref{int} and only the second term of the integral (which is also an integral) remains. Using the trapezoidal rule the integral term takes the form $\int_{a}^{b} f(x) \,dx\ \approx (b-a)\frac{f(a) + f(b)}{2}$. Assuming the shock front to be infinitesimally thin, we have $ b-a \approx 0 $. Thus eqn \ref{int} takes a simple form

\begin{align}
\frac{\partial_{0}T^{10}}{T^{11}} + \partial_{1}\Big(\log(T^{11} e^{2 \Lambda(r)})\Big) = 0 \label{2nd}. 
\end{align}

The two redundant equations are
\begin{align*}
      \nabla_{2}T^{22} = 0 \\
      \nabla_{3}T^{33} = 0.
\end{align*}

Similarly, we can find the simplified form of particle conservation equation from eqn (5) and is given by
\begin{align}
    \nabla_{0}(nu^{0}) + \nabla_{1}(nu^{1}) = 0 \nonumber \\
   \Rightarrow \partial_{0}(nu^{0}) + \partial_{1}(nu^{1}) + nu^{1}(\phi'(r) + \Lambda'(r)) =0 \nonumber \\
    \Rightarrow \frac{\partial_{0}(nu^{0})}{(nu^{1})} + \partial_{1}\Big(\log(nu^{1} e^{ \phi(r) + \Lambda(r)})\Big) = 0. \label{3rd}
\end{align}

Having derived the explicit form of the conservation equation for a curved ST, we next find the jump condition and Taub adiabat equation.
Considering the shock surface as a hypersurface, we will write the jump conditions taking shock front as our frame of reference. For SL shocks, the equation of hypersurface will be $r = constant$ and for TL shocks, $t = constant$. 
Denoting the RHS of the shock discontinuity (or upstream) as “a” and the LHS of the shock-discontinuity as “b” (shown in fig \ref{sides}), the difference of a thermodynamic quantity (say Z) across the front is given by $[Z] \equiv Z_{a}$-$Z_{b}$. Using eqn \ref{1st},\ref{2nd} and \ref{3rd}, we can write the jump conditions across the discontinuity for SL and TL shock waves as,

\begin{itemize}
    \item SL shocks 
    
    The energy-flux jump condition is 
    \begin{align}
      [T^{01}e^{(3 \phi + \Lambda)}] & = 0 \nonumber \\
      \Rightarrow w_{a}\gamma_{ga}^2 v_{a} e^{(3 \phi_a + \Lambda_a)} & = w_{b}\gamma_{gb}^2 v_{b} e^{(3 \phi_b + \Lambda_b)}.\label{grsl1}
      \end{align}
      \par The momentum-flux conservation jump condition becomes
      \begin{align}
      [T^{11}e^{2\Lambda}] & = 0 \nonumber \\
      \Rightarrow w_{a}\gamma_{ga}^2 v_{a}^2 e^{2 \Lambda_a} + {p_a} & =  w_{b}\gamma_{gb}^2 v_{b}^2 e^{2 \Lambda_b} + {p_b}.\label{grsl2}
      \end{align}
      \par The particle-flux conservation is given by \\
      \begin{align}
      [nu^{1} e^{(\phi + \Lambda)}] &= 0 \nonumber \\     
      \Rightarrow n_a\gamma_{ga}v_{a}e^{(\phi_a + \Lambda_a)} &=  n_b\gamma_{gb}v_{b}e^{(\phi_b + \Lambda_b)} =j. \label{grsl3}      \end{align}
    
    Using eqn \ref{grsl2} and \ref{grsl3} the particle-current density (j) is given by
    \begin{align}
        j^2 & = \frac{(p_b - p_a)}{(w_a V_a^2 e^{(\Lambda_a - \phi_a)} - w_b V_b^2 e^{(\Lambda_b - \phi_b)})} \label{grslj2}
    \end{align}
    \par where, $V_a=\frac{1}{n_a}$ and $V_b=\frac{1}{n_b}$. Using eqn \ref{grsl1}, \ref{grsl2} and \ref{grslj2} we can derive the Taub adiabat (TA) (or combustion adiabat (CA) if the equation of state (EoS) of side (or phase) "a" is different from that of side "b"). TA/CA is an equation relating the thermodynamic variables across the shock having no terms involving matter velocities. 
    
    For SL GR shock with $\mu = \frac{w}{n^2}$, it takes the form
    
    \begin{align}
      \frac{(p_b - p_a)\left[\mu_b^2V_b^2e^{2(\Lambda_b + \phi_b)} - \mu_a^2V_a^2e^{2(\Lambda_a + \phi_a)}\right]}{(\mu_b^2 e^{(3 \phi_b + \Lambda_b)} - \mu_a^2 e^{(3 \phi_a + \Lambda_a)})}\nonumber\\
     = (\mu_b V_b e^{(\Lambda_b - \phi_b)} - \mu_aV_a e^{(\Lambda_a - \phi_a)}). \label{tasl}
    \end{align}

The matter velocities of the two phases in terms of the thermodynamic variables can be solved from eqn \ref{grsl1}, \ref{grsl2} and are given by 
\begin{align}
    {v_a} &= \sqrt{\frac{a_{11} + w_a b_{11}}{c_{11}}} \label{v1}\\
    {v_b} &= \frac{{v_a}B_1(A_1^4 w_a + A_2^4 w_b  - b_{11})}{2 A_1^3 A_2 B_2[p_a + \epsilon_b]} \label{v2}
\end{align}
where, we have defined 
 \begin{align*}
     A_1 & = e^{\phi_a} \text{,   }
     A_2  = e^{\phi_b} \text{,   }
     B_1  = e^{\Lambda_a} \text{,   }
     B_2  = e^{\Lambda_b}\\
     w_a & = (p_a + \epsilon_a) \text{,   }
      w_b  = (p_b + \epsilon_b)\\
     a_{11} &= A_1^4 w_a^2 - A_2^4(p_a(p_b + 2 \epsilon_a - \epsilon_b) \\& + 2 p_b \epsilon_b - p_b \epsilon_a + \epsilon_a \epsilon_b)\\
     b_{11} &= \sqrt{w_b^2 A_2^8 - w_a^2 A_1^8 -  2 A_1^4 a_{11}}\\
     c_{11} &= 2 \frac{B_1^2}{A_1^2} A_2^4(p_b +\epsilon_a)(\epsilon_a -\epsilon_b). \\
 \end{align*}
     
    \item TL shocks
    
     \par The energy-density conservation is
      \begin{align}
      [T^{00}] & = 0 \nonumber\\
       \Rightarrow w_a\gamma_{ga}^2 - \frac{p_a}{e^{2\phi_a}} & = w_b\gamma_{gb}^2 - \frac{p_b}{e^{2\phi_b}}. \label{grtl1}
      \end{align}
    \par The momentum-density conservation becomes
    \begin{align}
      [T^{10}] & = 0 \nonumber\\
      \Rightarrow w_a\gamma_{ga}^2 v_{ra} & = w_b\gamma_{gb}^2 v_{rb} \label{grtl2} 
      \end{align}

      \par The particle number density conservation is \\
     \begin{align}
      [nu^{0}] & = 0 \nonumber \\
      \Rightarrow n_a\gamma_{ga} &=  n_b\gamma_{gb} = j \label{grtl3}
    \end{align}
    \par where j$^2$ is given by
    \begin{align}
         j^2 & = (\frac{p_a}{e^{2\phi_a}}-\frac{p_b}{e^{2\phi_b}}) \frac{1}{(w_aV_a^{2}-w_bV_b^{2})} \label{grtlj2}
    \end{align}
\end{itemize}

Similarly, using eqn \ref{grtl1},\ref{grtl2},\ref{grtlj2} one can write TA/CA for TL shocks as

\begin{align}
   &\left[\frac{p_b}{e^{2\phi_b}} - \frac{p_a}{e^{2\phi_a}}\right]\left[ \frac{\mu_b^2V_b^2e^{2\phi_b}}{e^{2\Lambda_b}} - \frac{\mu_a^2V_a^2e^{2\phi_a}}{e^{2\Lambda_a}}\right]\nonumber\\
    &= (\mu_bV_b - \mu_aV_a)\left[ \frac{\mu_b^2}{e^{2\Lambda_b}} -  \frac{\mu_a^2}{e^{2\Lambda_a}}\right] \label{tatl}. 
\end{align}

The matter velocities on either side of the discontinuity is given by

\begin{align}
    {v_a} &= \sqrt{\frac{A_1^{2}(a_{21} - b_{21})}{2c_{21}}} \\
    {v_b} &= \frac{{v_a}A_2^{2}(B_2^{2} w_a + B_1^{2} w_b  + \frac{b_{21}}{w_a A_1A_2} )}{2B_2^{2}[A_1^{2}p_b + A_2^{2}\epsilon_a]}
\end{align}
 where, 
 \begin{align*}
     w_a & = (p_a + \epsilon_a) \text{,}
      w_b  = (p_b + \epsilon_b)\\
     a_{21} &= A_1^{2}A_2^{2}\big[B_2^{2} w_a^{2} - B_1^{2}( \epsilon_b - p_b)( \epsilon_a - p_a)] \\
     &- 2B_1^{2}(A_2^{2}p_a\epsilon_a +A_1^{2}p_b\epsilon_b)\\
     b_{21} &= w_aA_1A_2\sqrt{A_1^{2}A_2^{2}(B_1^{4}w_b^{2}-B_2^{4}w_a^{2}) + 2B_2^{2}a_{21}}\\
     c_{21} &= B_1^{4}(A_2^{2}p_a + A_1^{2}\epsilon_b)(A_2^{2}p_a - A_1^{2}p_b)   . \\
 \end{align*}

Equations of the GR shocks reduce to the corresponding SR shocks (both for SL and TL) if the metric potentials becomes zero ($ \Lambda_a = \phi_a = \Lambda_b = \phi_b = 0 $). The SR TA/CA equation becomes 
\begin{equation*}
     \mu_b^2 - \mu_a^2 = (p_b - p_a)(\mu_b V_b + \mu_a V_a). \\
\end{equation*}

It is also interesting to note that for the GR shocks the TA/CA equations are different for the SL and TL shocks, unlike in the SR case where they are the same for both.

\section{Weak Shock waves}

Before studying shock properties in general, it is useful to check how the thermodynamic quantities vary across the shock if the discontinuity is not very large. This type of discontinuity is usually known as weak shock, and it gives useful insight into the entropy change across the shock discontinuity.
Weak shocks signify that the state of medium "b" (near the shock discontinuity) can be calculated using Taylor expansion around the variable of the phase "a".

\subsection{Special Relativistic weak SL/TL shocks}
We expand the SR TA/CA equation, i.e $\mu_b^2$ and $\mu_b V_b$ keeping terms up to third order in $p_b-p_a$, but only up to first order in $s_b - s_a$ \cite{thorne}. Starting from the TA equation of SR shocks, we have

\begin{align}
    \mu_b^2 - \mu_a^2 = (p_b - p_a)(\mu_b V_b + \mu_a V_a) \label{srta}
\end{align}
Expanding as prescribed, we have
 \begin{align*}
    &\mu_b^2 = \mu_a^2 + (p_b - p_a)\diffp {(\mu^2)}p[s,1] + \\
    &(s_b - s_a)\diffp {(\mu^2)}s[p,1] 
     + \frac{1}{2}(p_b - p_a)^2\diffp[2]{(\mu^2)}p[s,1] +\\
     &(p_b - p_a)(s_b - s_a)\diffp[1,1]{(\mu^2)}{p,s} 
      + \frac{1}{6}(p_b - p_a)^3\diffp[3]{(\mu^2)}p[s,1] + \\ 
      &\frac{1}{2}(p_b - p_a)^2(s_b - s_a)\diffp[2,1]{(\mu^2)}{p,s} +...
\end{align*} 

Using the 1st law of thermodynamics, $d(\mu^2) = 2\mu d\mu = 2\mu Vdp + 2 \mu Tds $, in above equation, it becomes

\begin{align}
    \mu_b^2 &= \mu_a^2 + 2 \Bigg[ (p_b - p_a) \mu_a V_a + (s_b - s_a) \mu_a T_a \nonumber \\
     & + \frac{1}{2}(p_b - p_a)^2\diffp{(\mu V)}p[s,1] + (p_b - p_a)(s_b - s_a)\diffp{(\mu V)}s \nonumber \\
     & + \frac{1}{6}(p_b - p_a)^3\diffp[2]{(\mu V)}p + \frac{1}{2}(p_b - p_a)^2(s_b - s_a)\diffp[1,1]{(\mu V)}{p,s}\Bigg] \nonumber\\
     & + O[(p_b - p_a)^4] + ...    \label{mub}
\end{align} 

Similarly, $\mu_b V_b$ can be expanded as

\begin{align}
   & \mu_b V_b = \mu_a V_a + (p_b - p_a)\diffp {(\mu V)}p[s,1] +  \nonumber\\ 
    &(s_b - s_a)\diffp {(\mu V)}s[p,1] 
      + \frac{1}{2}(p_b - p_a)^2\diffp[2]{(\mu V)}p[s,1] +\nonumber \\
     & (p_b - p_a)(s_b - s_a)\diffp[1,1]{(\mu V)}{p,s}
      + ... \label{muv}
\end{align} 

Using eqn \ref{mub} and eqn \ref{muv} in eqn \ref{srta}, and solving for $s_b - s_a$ we have

\begin{align}
    &s_b - s_a  = \frac{\frac{1}{12 \mu_a T_a} \diffp[2]{(\mu V)}p(p_b - p_a)^3}{\big( 1 + \frac{(p_b - p_a)}{2 \mu_a T_a }\diffp{(\mu V)}s \big)} + ...\nonumber\\
    &=\frac{1}{12 \mu_a T_a} \diffp[2]{(\mu V)}p(p_b - p_a)^3 + O[(p_b - p_a)^4] +... \label{ssr}
\end{align}

\begin{figure}
\centering
\hspace{0.5cm}
  \includegraphics[scale=0.8]{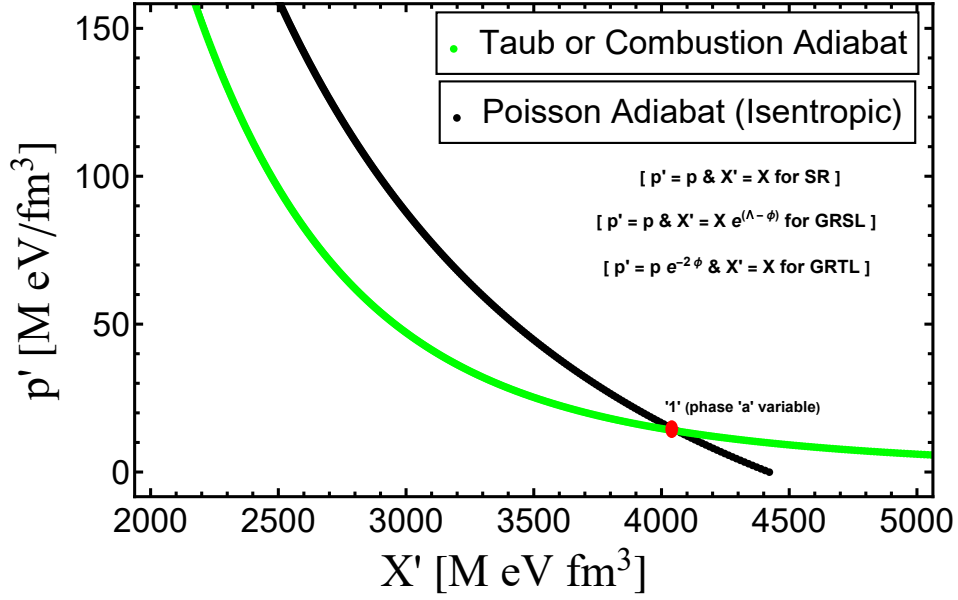}
  \caption{Under weak shock waves, the plots of combustion adiabat and Poisson adiabat overlap near point "1" which means shock adiabat or combustion adiabat will be isentropic near point"1". }
  \label{ws}
\end{figure}

Since in weak shock waves, the variation in thermodynamic quantities is small, so from eqn \ref{ssr} we find that weak shock wave phenomenons are isentropic. As shown in fig \ref{ws}, near point "1" we have an isentropic process where the Poisson adiabat and TA/CA adiabat intersect.  

\subsection{General Relativistic weak SL shocks}
Similarly, we find the entropy change in the GRSL and GRTL shocks. As the TA/CA equation is different, we find their expression separately. Writing eqn \ref{tasl} as,

\begin{align}
    &(p_b - p_a)\left(\mu_b V_b h_b -\mu_a V_a h_a\right)\left(\mu_b V_b h_b + \mu_a V_a h_a\right) \nonumber \\
    & = (\mu_b V_b f_b - \mu_a V_a f_a)\left(\mu^2_b g_b -\mu^2_a g_a \right) \label{grslw1} 
\end{align}
where, $g_a = A_1^3 B_1$, $g_b = A_2^3 B_2$, $f_a = \frac{B_1}{A_1}$, $f_b = \frac{B_2}{A_2}$, $h_a = \sqrt{g_a f_a}$ and  $h_b = \sqrt{g_b f_b}$ .
Considering the metric potentials to be varying slowly for the system, we can expand the above equation using Taylor expansion as

\begin{align}
   & \mu_b V_b h_b = \mu_a V_a h_a + h_a (p_b - p_a)\diffp {(\mu V)}p[s,1]  \nonumber \\
    &+  h_a [(s_b - s_a)\diffp {(\mu V)}s[p,1]
    + (p_b - p_a)(s_b - s_a)\diffp[1,1]{(\mu V)}{p,s} \nonumber \\
    &+ \frac{1}{2}(p_b - p_a)^2\diffp[2]{(\mu V)}p[s,1] + ...]\label{grslw2}
\end{align} 

and 

\begin{align}
    &\mu_b V_b f_b = \mu_a V_a f_a + f_a (p_b - p_a)\diffp {(\mu V)}p[s,1]  \nonumber \\
    &+  f_a [(s_b - s_a)\diffp {(\mu V)}s[p,1]
    + (p_b - p_a)(s_b - s_a)\diffp[1,1]{(\mu V)}{p,s} \nonumber \\
    &+ \frac{1}{2}(p_b - p_a)^2\diffp[2]{(\mu V)}p[s,1] + ...]\label{grslw3}
\end{align} 

Similarly, $\mu^2_b g_b$ can be expanded with the use of 1st Law of thermodynamics as,

\begin{align}
    &\mu_b^2 g_b = \mu_a^2 g_a + \frac{g_a}{3} \Bigg[ 6(p_b - p_a) \mu_a V_a + 6(s_b - s_a) \mu_a T_a \nonumber \\
    &  + 3(p_b - p_a)^2\diffp{(\mu V)}p[s,1] + 6(p_b - p_a)(s_b - s_a)\diffp{(\mu V)}s \nonumber \\
     & + (p_b - p_a)^3\diffp[2]{(\mu V)}p + 3(p_b - p_a)^2(s_b - s_a)\diffp[1,1]{(\mu V)}{p,s}\Bigg] \nonumber\\
     &+ O[(p_b - p_a)^4] + ... \label{grslw4}
\end{align} 

Using eqn \ref{grslw2},\ref{grslw3},\ref{grslw4} in eqn \ref{grslw1}, the change in entropy for weak GRSL shocks is given by

\begin{align}
   & s_b - s_a  = \frac{\frac{1}{12 \mu_a T_a} \diffp[2]{(\mu V)}p(p_b - p_a)^3}{\big( 1 + \frac{(p_b - p_a)}{2 \mu_a T_a }\diffp{(\mu V)}s \big)} + ...\nonumber\\
   & =\frac{1}{12 \mu_a T_a} \diffp[2]{(\mu V)}p(p_b - p_a)^3 + O[(p_b - p_a)^4] +... \label{sgrsl}
\end{align}

Interestingly, the entropy change expression for weak SR shock and GRSL comes out to be the same.

\subsection{General Relativistic weak TL shocks}
Under similar considerations, the entropy change for the GRTL shock can be derived from the GRTL TA equation (eqn \ref{tatl})
\begin{align}
    &(\frac{p_b g_b^2}{f_b^2} - \frac{p_a g_a^2}{f_a^2})\left(\mu_b V_b f_b -\mu_a V_a f_a\right)\left(\mu_b V_b f_b + \mu_a V_a f_a\right) \nonumber \\
     &= (\mu_b V_b - \mu_a V_a)\left(\mu^2_b g^2_b -\mu^2_a g^2_a \right) \label{grtlw1}
\end{align}
where,  $g_a = \frac{1}{e^{\Lambda_a}}$, $g_b = \frac{1}{e^{\Lambda_b}}$, $f_a = \frac{e^{\phi_a}}{e^{\Lambda_a}}$, $f_b = \frac{e^{\phi_b}}{e^{\Lambda_b}}$.

Expanding the above equation using Taylor expansion around medium "a" variables, 
and following the same procedure, we have





\begin{align}
    s_b - s_a &= \frac{A\bigg(2\mu_a V_a + (p2-p1)\diffp{(\mu V)}p\bigg) + B\diffp[2]{(\mu V)}p}{6\bigg(2 f^2_b g^2_a T_a \mu_a + C \diffp{(\mu V)}s \bigg) + D\diffp[1,1]{(\mu V)}{p,s}} +...\label{stl}
\end{align}
where, A, B, C, D defined as

\begin{align*}
    A &= 6 p_b (f_b g_a - f_a g_b)(f_b g_a + f_a g_b)\\
    B &= (p_b - p_a)^2(f^2_b g^2_a(2p_b + p_a) - 3 f^2_a g^2_b p_b)\\
    C &= (f^2_b g^2_a(2p_b - p_a) -  f^2_a g^2_b p_b) \\
    D &= p_b (p_b - p_a) (f_b g_a - f_a g_b)(f_b g_a + f_a g_b)
\end{align*}

It can be seen that $f$'s and $g$'s are the function of metric potentials, and to have the knowledge of entropy change for weak GR TL shock, one needs to have some definite metric potential of an astrophysical system.

\section{Strong Shock Waves and Chapman-Jouguet Point}
\begin{figure}
\centering
\hspace{0.5cm}
  \includegraphics[scale=0.8]{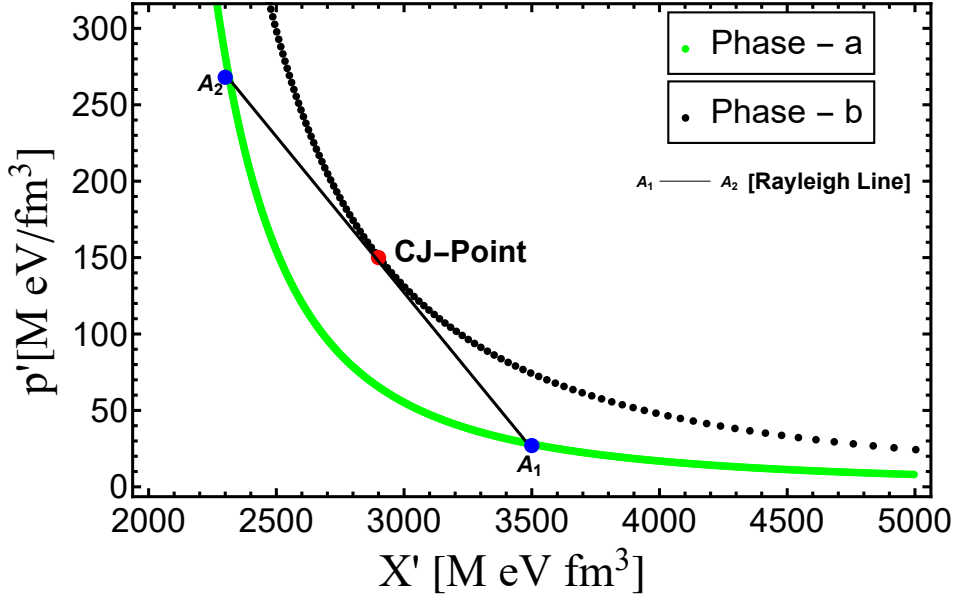}
  \caption{Plot showing the CJ-point, which is the tangent on combustion adiabat from a shock adiabat curve where the line "A1-A2" is known as the Rayleigh line.} \label{cjfig}
\end{figure}

Shock waves with arbitrary strength are considered to be strong shocks. For strong shock waves, it can be shown that nowhere on the TA can the "chord" (straight line connecting medium "a" variable to medium "b") become tangent to the adiabat if combustion does not happen. But if combustion occurs (due to shock) waves, then a tangent can exist, and the tangent point is known as the CJ point, and the line connecting the two points is called the Rayleigh line (RL), as shown in fig \ref{cjfig} . 

\subsection{Special Relativistic Strong Shock wave and CJ point} 
The TA/CA equation in the SR case is the same for both SL and TL shocks. Still, we get different CJ points because $j^2$ is different. First, we construct a differential equation for the squared baryon flux ($j^2$) as a function of entropy ($s_b$), and then we calculate the CJ point. Differentiating the TA equation (eqn \ref{srta}) by having medium "a" variables fix, we have
\begin{align*}
    2\mu_b \dd \mu_b &= (\mu_a V_a + \mu_b V_b)\dd p_b + (p_b - p_a)\dd (\mu_b V_b)
\end{align*}
Using the 1st law of thermodynamics ($\dd \mu_b = V_b\dd p_b + T_b \dd s_b$) we have
\begin{align}
    2 \mu_b T_b \dd s_b &= (\mu_a V_a - \mu_b V_b)\dd p_b + (p_b - p_a)\dd (\mu_b V_b) \label{dsrta}
\end{align}
Next, we need to find $j^2$ for both SL and TL cases to get the CJ point.

\subsubsection{For SL Shocks}
The squared baryon flux ($j^2$) is given by,
\begin{align}
    j^2 &= \frac{p_a - p_b}{\mu_b V_b - \mu_a V_a} \nonumber  \\
    \Rightarrow \dd j^2 &= \frac{(\mu_a V_a - \mu_b V_b)\dd p_b + (p_b - p_a)\dd (\mu_b V_b)}{(\mu_b V_b - \mu_a V_a)^2} \label{dj2}
\end{align}
Using equation eqn \ref{dsrta} in above equation, we get

\begin{align}
    \dd j^2 &= \frac{2 \mu_b T_b \dd s_b}{(\mu_b V_b - \mu_a V_a)^2} 
\end{align}

Considering there exists a point at which the chord is tangent to the adiabat, and because the slope of the chord is $-j^2$, we must have

\begin{align}
    \dv {j^2}{p_b} &= \dv{s_b}{p_b} = 0 \text{  on the CA at CJ point.} \label{cj1}
\end{align}

From eqn \ref{dj2}, we can obtain 

\begin{align*}
    \dv {j^2}{p_b} &= \frac{1 + j^2 \dv {(\mu_b V_b)}{p_b}}{(\mu_b V_b - \mu_a V_a)} 
\end{align*}

For this to be zero, we have the condition 
\begin{align*}
    1 + j^2 \diff{\mu_b V_b}{p_b}{s_b} &= 0 \\
    \Rightarrow 1 &= -(\frac{u^2_b}{V^2_b})\diff {(\mu_b V_b)}{p_b}{s_b}
\end{align*}
where, $\diff {(\mu_b V_b)}{p_b}{s_b}$ = $\frac{-V^2_b}{u^2_{bs}}$, $u_{bs}$ = 4-velocity of sound in medium "b".

So, from the above equation if "chord = tangent" then at that point,
\begin{align}
    u^2_b &= u^2_{bs} \text{ at CJ point }.  \label{cj2} 
\end{align}

The above equation implies that, at CJ point, the downstream matter moves sonically. 

\subsubsection{For TL Shocks}
For the TL shocks, the squared baryon flux ($j^2$) is given by
\begin{align}
    j^2 &= \frac{p_b - p_a}{\mu_b V_b - \mu_a V_a} \nonumber\\
    \Rightarrow \dd j^2 &= \frac{(\mu_b V_b - \mu_a V_a)\dd p_b - (p_b - p_a)\dd (\mu_b V_b)}{(\mu_b V_b - \mu_a V_a)^2} \label{dj2tl} 
\end{align}
Using the equation in eqn \ref{dsrta}, we get

\begin{align}
    \dd j^2 &= -\frac{2 \mu_b T_b \dd s_b}{(\mu_b V_b - \mu_a V_a)^2} 
\end{align}

At the CJ point we muxt have

\begin{align}
    \dv {j^2}{p_b} &= \dv{s_b}{p_b} = 0 \label{cj1tl}
\end{align}

From eqn \ref{dj2tl}, we obtain

\begin{align*}
    \dv {j^2}{p_b} &= \frac{1 - j^2 \dv {(\mu_b V_b)}{p_b}}{(\mu_b V_b - \mu_a V_a)} 
\end{align*}
and for it to be zero, we must have
\begin{align*}
    1 - j^2 \diff{\mu_b V_b}{p_b}{s_b} &= 0 \\
    \Rightarrow 1 &= (\frac{u^2_b}{V^2_b})\diff {(\mu_b V_b)}{p_b}{s_b}
\end{align*}
where, $\diff {(\mu_b V_b)}{p_b}{s_b}$ = $\frac{-V^2_b}{u^2_{bs}-1}$, $u_{bs}$ = 4-velocity of sound in medium "b".

Therefore, for the TL shocks if "chord = tangent" then at CJ point,
\begin{align}
    u^2_b &= 1- u^2_{bs}   
\end{align}

\subsection{GR Strong Shocks and CJ point}
To analyze strong GR SL and TL shocks, we follow the same procedure that was followed for SR analysis. The differential equation for the squared baryon flux ($j^2$) as a function of entropy $s_b$ is first constructed, and then we find the CJ point.
\subsubsection{For GR SL Shocks}
Differentiating eqn \ref{tasl} by having medium "a" variables (and assuming metric potentials are also not changing) fixed, we have
\begin{align*}
    & 2\mu_b e^{(3 \phi_b + \Lambda_b)} (\mu_b V_b e^{(\Lambda_b - \phi_b)} - \mu_a V_a e^{(\Lambda_a - \phi_a)})^2 \dd \mu_b \\
    &= (\mu_b V_b e^{(\Lambda_b - \phi_b)} - \mu_a V_a e^{(\Lambda_a - \phi_a)})\Bigg[\dd p_b\Bigg(\frac{\mu_b^2V_b^2e^{2\Lambda_b}}{e^{-2\phi_b}} \nonumber \\
    &- \frac{\mu_a^2V_a^2e^{2\Lambda_a}}{e^{-2\phi_a}} \Bigg) +(p_b - p_a)\frac{2\mu_b V_b e^{2\Lambda_b} \dd (\mu_b V_b) }{e^{-2\phi_b}} \Bigg]\nonumber\\
    & - (p_b - p_a)\Bigg(\frac{\mu_b^2V_b^2e^{2\Lambda_b}}{e^{-2\phi_b}} - \frac{\mu_a^2V_a^2e^{2\Lambda_a}}{e^{-2\phi_a}}\Bigg)e^{(\Lambda_b-\phi_b)}\dd (\mu_b V_b)
\end{align*}

and from eqn \ref{grslj2} we can get,

\begin{align}
    \dd (\mu_b V_b) &= \frac{\dd(j^2)(\mu_b V_b e^{(\Lambda_b-\phi_b)} - \mu_a V_a e^{(\Lambda_a-\phi_a)})^2}{{(p_b - p_a)e^{(\Lambda_b-\phi_b)}}} +\nonumber \\
    & \frac{\dd(p_b)(\mu_b V_b e^{(\Lambda_b-\phi_b)} - \mu_a V_a e^{(\Lambda_a-\phi_a)})}{{(p_b - p_a)e^{(\Lambda_b-\phi_b)}}} \label{grsldj2}
\end{align}
Using 1st law of thermodynamics and $\dd (\mu_b V_b)$, we have
\begin{align*}
    \dd j^2 &= \frac{2 A_2^3 B_2 T_b \mu_b ds_b}{A_2^2 B_2^2 \mu_b^2 V_b^2 + \frac{B_1 \mu_a V_a(A_1^3 B_1 V_a \mu_a - 2 A_2^3 B_2 V_b \mu_b)}{A_1}}
\end{align*}

If the chord becomes tangent then it must follow the conditions,
\begin{align}
    \dv {j^2}{p_b} &= \dv{s_b}{p_b} = 0 \text{  on the CA at CJ point.} 
\end{align}
From eqn \ref{grsldj2}, we have 

\begin{align}
    \dv {j^2}{p_b} &= \frac{-1 - j^2 e^{(\Lambda_b - \phi_b)}\diff {(\mu_b V_b)}{p_b}{s_b}}{(\mu_b V_b e^{(\Lambda_b-\phi_b)} - \mu_a V_a e^{(\Lambda_a - \phi_a)})} \label{dj2grsl}
\end{align}

Using eqn \ref{grsl3} for the $j^2 = \frac{u_b^2 (e^{2(\Lambda_b + \phi_b)})}{V_b^2} $, $ \diff {(\mu_b V_b)}{p_b}{s_b} = V^2_b (1 - \frac{1}{v^2_{sb}} )$ and the expression for 4-velocity in GR SL case, we  can write the 4-velocity of sound ($u_{sb}$) relative to fluid in terms of 3-velocity ($v_{sb}$) of sound in medium "b" as 
\begin{align*}
    u^2_{sb} &= \frac{v^2_{sb}}{(e^{2\phi_b} - v^2_{sb} e^{2 \Lambda_b})}
\end{align*}

For eqn \ref{dj2grsl} to be zero, we must have,
\begin{align}
    -1 - j^2 e^{(\Lambda_b - \phi_b)}\diff {(\mu_b V_b)}{p_b}{s_b} = 0\nonumber\\
    \Rightarrow u^2_b = \frac{u^2_{sb} e^{\phi_b}}{[e^{3\Lambda_b}(1- u^2_{sb}(e^{2\phi_b}-e^{2\Lambda_b}))]} \text{  at CJ point} \label{grslcj}
\end{align}

Therefore, for GR SL shocks, the CJ point is not explicitly a sonic point.

\subsubsection{For GR TL Shocks}
 
Similarly, differentiating the TL TA/CA equation we have
\begin{align*}
    &\frac{ 2\mu_b (\mu_b V_b - \mu_a V_a)^2 \dd \mu_b}{e^{2\Lambda_b}} = (\mu_b V_b - \mu_a V_a)\Bigg[\frac{\dd p_b}{e^{2 \phi_b}}\Bigg(\frac{\mu_b^2V_b^2e^{2\phi_b}}{e^{2\Lambda_b}} \nonumber \\
    &- \frac{\mu_a^2V_a^2e^{2\phi_a}}{e^{2\Lambda_a}} \Bigg) + \Bigg(\frac{p_b}{e^{2\phi_b}} - \frac{p_a}{e^{2\phi_a}}\Bigg)\frac{2\mu_b V_b e^{2\phi_b} \dd (\mu_b V_b) }{e^{2\Lambda_b}} \Bigg]\nonumber\\
    & - \Bigg(\frac{p_b}{e^{2\phi_b}} - \frac{p_a}{e^{2\phi_a}}\Bigg) \Bigg(\frac{\mu_b^2V_b^2e^{2\phi_b}}{e^{2\Lambda_b}} - \frac{\mu_a^2V_a^2e^{2\phi_a}}{e^{2\Lambda_a}}\Bigg)\dd (\mu_b V_b)
\end{align*}
Using 1st law of thermodynamics and value of $\dd (\mu_b V_b)$ becomes
\begin{align}
    \dd (\mu_b V_b) &= \frac{\frac{(\mu_b V_b - \mu_a V_a)\dd p_b}{e^{2 \phi_b}} - (\mu_b V_b - \mu_a V_a)^2 \dd (j^2)}{\bigg(\frac{p_b}{e^{2\phi_b}} - \frac{p_a}{e^{2\phi_a}}\bigg)} \label{dj2grtl}
\end{align}
From the above equation, we have
\begin{align*}
    \dd j^2 &= -\frac{2 e^{2 \Lambda_a} T_b \mu_b \dd s_b}{( e^{2(\phi_a+\Lambda_b)}\mu^2_a V^2_a + \mu_b V_b e^{2(\phi_b + \Lambda_a)} (\mu_b V_b - 2 \mu_a V_a))}
\end{align*}

The condition for the chord to be the tangent is
\begin{align}
    \dv {j^2}{p_b} &= \dv{s_b}{p_b} = 0 \text{  on the CA at CJ point.}
    \end{align}
 
From eqn \ref{dj2grtl}, we have 
\begin{align}
    \dv {j^2}{p_b} &= \frac{\frac{1}{e^{2\phi_b}} - j^2 \diff {(\mu_b V_b)}{p_b}{s_b}}{(\mu_b V_b - \mu_a V_a)}\label{cjtl}
\end{align}
where, $ \diff {(\mu_b V_b)}{p_b}{s_b} = V^2_b (1 - \frac{1}{v^2_{sb}} ) $. The expression for the 4-velocity of sound ($u_{sb}$) relative to fluid in terms of 3-velocity ($v_{sb}$) of sound in medium "b" is 
\begin{align*}
    u^2_{sb} &= \frac{1}{(e^{2\phi_b} - v^2_{sb} e^{2 \Lambda_b})}.
\end{align*}

Therefore, at CJ point from eqn \ref{cjtl} we must have,
\begin{align}
    \frac{1}{e^{2\phi_b}} - j^2 \diff {(\mu_b V_b)}{p_b}{s_b} = 0\nonumber\\
    \Rightarrow u^2_b = \frac{u^2_{sb}e^{2\phi_b}-1}{[e^{2\phi_b}( u^2_{sb}(e^{2\phi_b}-e^{2\Lambda_b})-1)]} \text{  at CJ point} \label{grtlcj}
\end{align}
Therefore, the downstream matter does not move sonically even for the GRTL shocks at the CJ point.

It is easy to check that equations of the GR shocks reduce to the corresponding SR shocks (both for SL and TL) if either the metric potentials becomes zero.
\begin{equation*}
     \Rightarrow \Lambda_a = \phi_a = \Lambda_b = \phi_b = 0 \\
     \end{equation*}
or if the metric potentials on either side of the front becomes equal, i.e. \\
 \begin{equation*}
     \Lambda_a = \Lambda_b = \Lambda; ~ \phi_a = \phi_b = \phi. \\
 \end{equation*}

\section{Neutron Stars as the system to study general relativistic shocks}

As stated earlier, one needs an astrophysical system to analyze the results obtained in the above sections.
We assume our astrophysical system to be a neutron star (NS). The metric described in section 2.1 is well suited to describe a non-rotating NS. Using an EoS and solving the so-called TOV equation \cite{tov}, we can determine the mass and radius of the star for a certain central energy density along with the variation of the pressure, energy density, and metric potential across the star. We assume that as the shock wave moves from the center of the star to the surface and is accompanied by conversion (or combustion/phase transition) of hadronic matter (HM) to quark matter (QM) \cite{p1,p2,p3,p4}. The PT happens till few kilometers from the centre of the star and stops at a distance where HM becomes more stable than the quark matter \cite{ritam-slow}. 
To describe HM we employ DD@BPS EoS \cite{eos} and for QM we employ MIT bag model having interactions \cite{chodos,alford,weissenborn}. The parameters used to describe the bag model are $B^{1/4}=160,a4=0.60$. As the PT happens, we ultimately have a star that has a quark core and hadronic outer surface, which we refer to as a hybrid star (HS).

\subsection{Weak Shock waves}
\begin{figure}
\centering
\hspace{0.5cm}
  \includegraphics[scale=0.8]{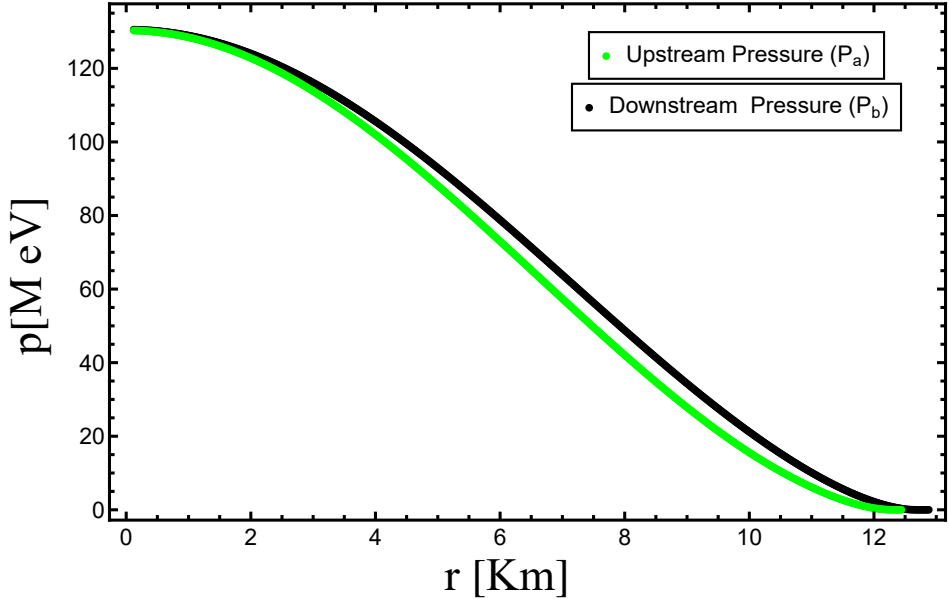}
  \caption{The plot of pressure vs. radius of an NS corresponding to HM EoS for upstream matter (Green) and QM EoS for downstream matter (Black) is shown in the figure.}\label{prs-vary}
\end{figure}

Once we generate a star and get the variation of the metric potential across it, we can calculate the entropy change due to shock propagation in the GR TL and GR SL shocks.
The plot of radius versus pressure of upstream matter (HM) and downstream matter (QM) shows that the difference in pressure is very small, which means that the entropy change for GR SL shocks is almost zero (fig \ref{prs-vary}). So, one can consider weak shock waves in NSs to be isentropic for GR SL shocks.

\begin{figure}
\centering
\hspace{0.5cm}
  \includegraphics[scale=0.8]{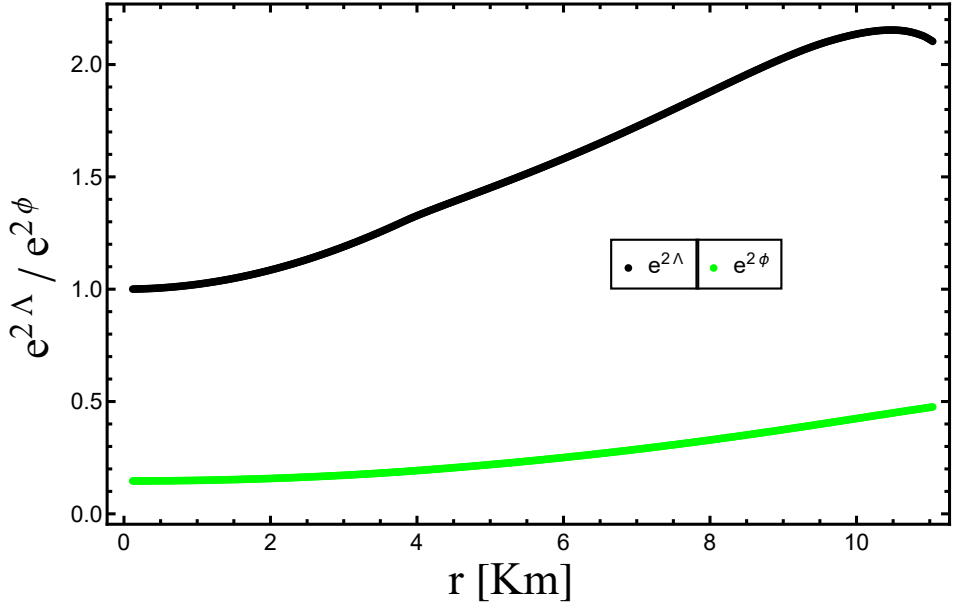}
  \caption{The plot of the metric potentials vs. radius corresponding to an HS is shown. The temporal component is marked with green, and the radial component is marked with black.}
  \label{f1}
\end{figure}

\begin{figure}
\centering
\hspace{0.5cm}
  \includegraphics[scale=0.8]{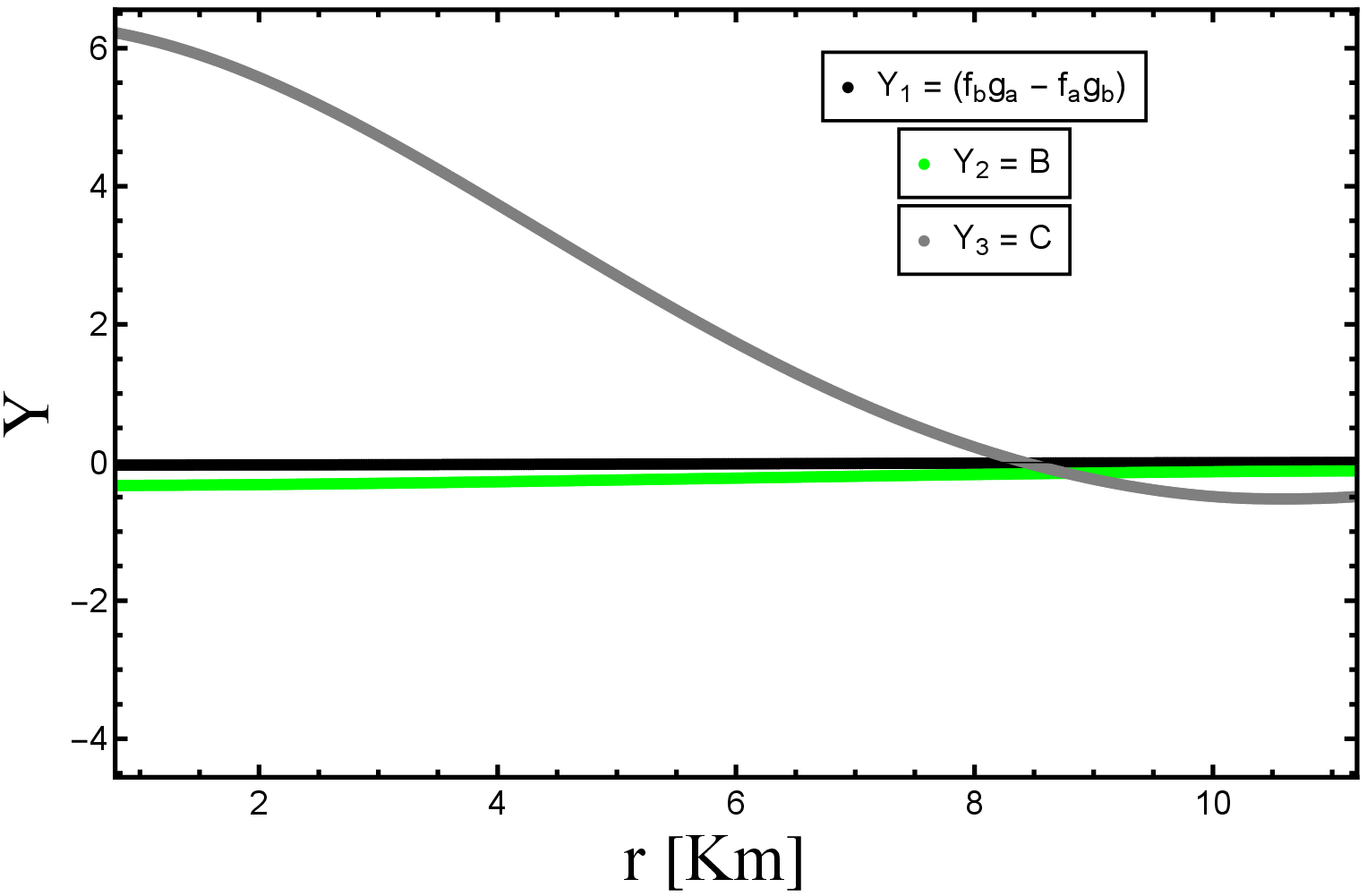}
  \caption{The plot is showing the potential factors $Y_1,Y_2,Y_3$ as a function of radius is illustrated in the figure.}\label{f7}
\end{figure}

The form of the expression of entropy for the GRTL shocks is different from the rest as it depends on the metric potentials. We have plotted the variation of metric potentials as a function of radius in fig \ref{f1}. As clear from the figure, both are slowly varying, and the spatial metric potential is always greater than the temporal metric potential.
Defining the metric potential difference factor as $Y_1 \equiv (f_bg_a - f_a g_b)$ we plot its variation with radius in fig \ref{f7}. 
From fig \ref{f7}, it is clear that $Y_1$ is zero throughout the star, which means A and D of eqn \ref{stl} become zero. We can also see the variation of B and C of eqn \ref{stl} with radius in the same figure. The variation of B with radius is small, but C varies considerably. However, as clear from the expression, it does not influence the equation much (eqn \ref{stl}), and the overall change of entropy with radius inside the NS for the GRTL case is also infinitesimal. Hence, even the weak GR TL shock wave is also isentropic in nature.     

\subsection{Downstream velocity at CJ point for GR shocks}
\begin{figure}
\centering
\hspace{0.5cm}
  \includegraphics[scale=0.8]{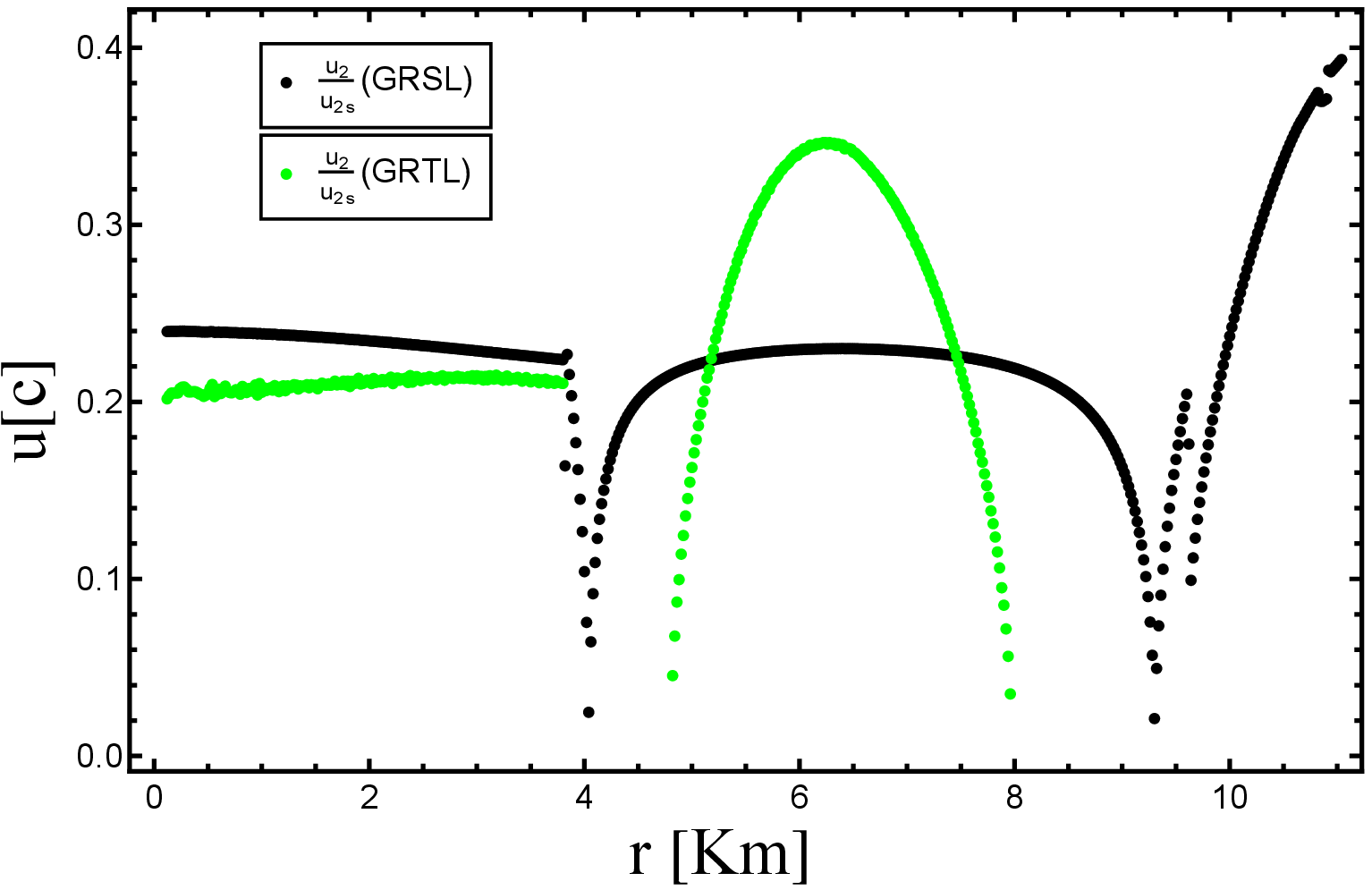}
  \caption{Velocity as a function of radius for the downstream matter at CJ point is plotted for both GRSL (Black), and GRTL (Green) shocks. The plot shows that near the core of the star, 4-velocity at CJ point for GRSL shock is more than that of GRTL shock; as we go towards the surface, GRTL 4-velocity exceeds; however, near the crust, GRSL 4-velocity remains finite whereas GRTL 4-velocity becomes imaginary or zero at CJ point.}
  \label{f2}
\end{figure}

The 4-velocity of the downstream matter is directly related to the local velocity of sound at the CJ point for SR shocks. 
But for GR shocks, we can interpret the 4-velocity clearly only after considering an astrophysical model (here, it is an NS) as they depend upon metric potentials. We have the variation of metric potentials with radius in fig \ref{f1} for downstream matter (QM). We can solve the eqn \ref{grslcj} and \ref{grtlcj} numerically to see the downstream matter velocity variation with respect to radius inside the NS and to check whether anywhere inside the star we can have a CJ point. Fig \ref{f2} shows the variation of the ratio of the velocity of downstream matter over the local velocity of sound as a function of radius for both GR SL and TL strong shocks at CJ point. 
As the ratio never reaches the value of $1$, we can interpret that the 4-velocity of downstream matter never reaches sonic speed anywhere inside the star for both SL and TL shocks. 

\subsection{Combustion through shock waves}

Among many problems where the GR shock treatment is necessary, one of them is the shock-induced PT in NSs.
As already stated, we assume that as the shock wave moves from the centre of the star to the surface, it is accompanied by a combustion (or phase transition) front, which converts HM to QM \cite{kamp,hana,dim}. The initial or upstream matter is assumed to be HM, and the final or downstream burnt matter is QM. Therefore, in this work, we study a process at the centre of an NS where HM is undergoing a deconfinement to QM due to the shock-induced burning.

In the present work, we focus only on GRSL CA and study the combustion of NS to hybrid star (HS) (the SR study has been done earlier \cite{m2,m3}). We study mainly SL shock because of the fact that in most astrophysical scenarios, the shock waves are considered to be SL. For TL shocks, the matter velocities sometimes become superluminous, which sometimes are interpreted as nonphysical in many scenarios. However, we have shown plots of GR TL shocks in the Appendix.

\begin{figure}
\centering
\hspace{0.5cm}
  \includegraphics[scale=0.8]{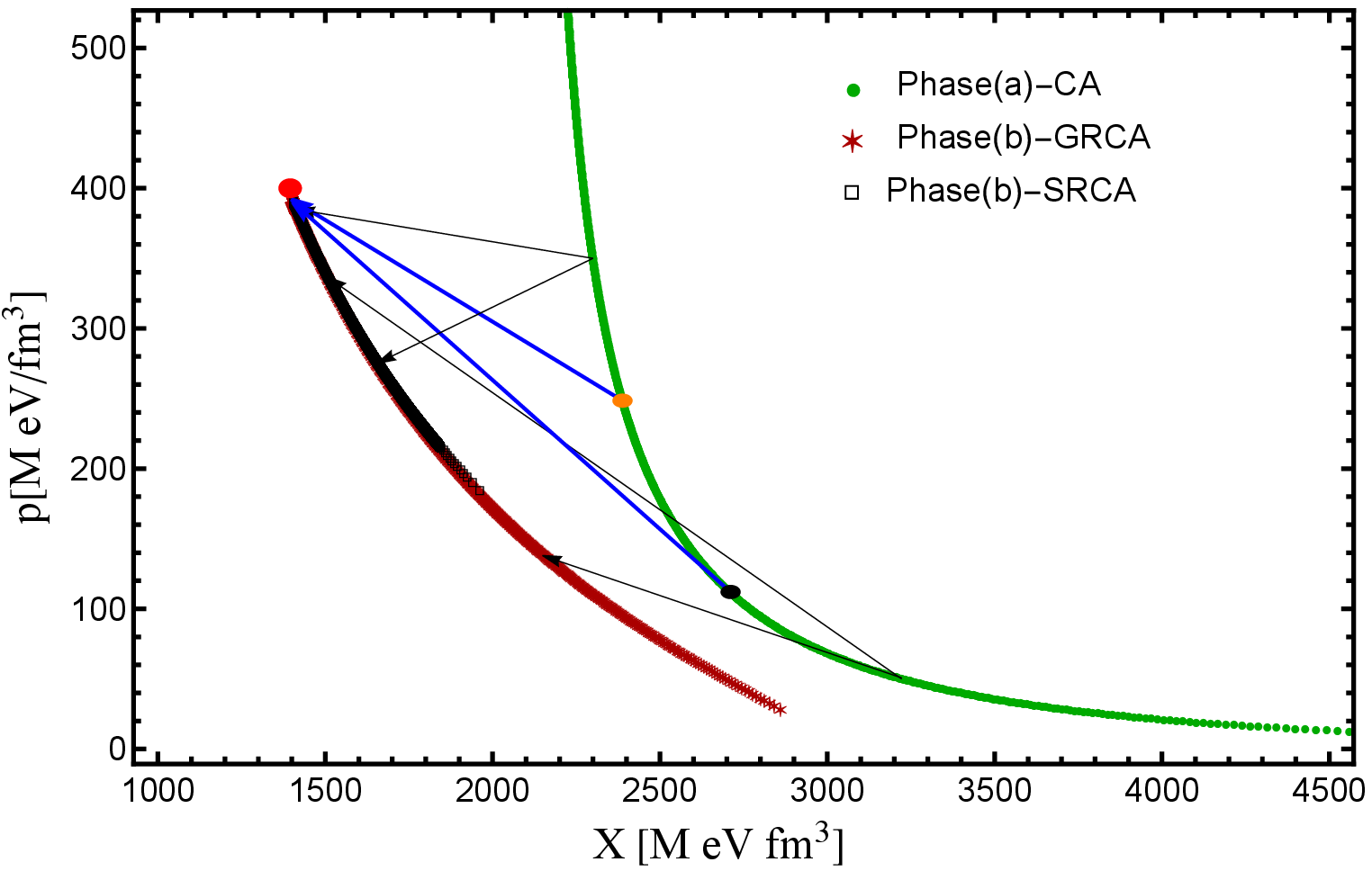}
  \caption{GR CA (SL) and SR CA (p versus X) curve show HM EoS (green-solid line with circle) with their corresponding burnt state QM (red-star). The red dot on QM EoS indicates the maximum pressure value corresponding to the orange dot on HM EoS. The arrows indicate the jump from HM EoS to its burnt state, and as the pressure becomes higher, the arrows indicate the retracing nature of the QM curves.}
  \label{f4}
\end{figure}

In our previous paper, SR CA was used to predict and constrain the maximum mass of the daughter HS which results from the combustion of parent NS \cite{}. In the present work, we extend the study and include GR corrections.
The initial state or the upstream quantities are the input (here, the HM EoS). Also, the final EoS of the burnt or downstream state is also known (the QM EoS). The CA is used to calculate the corresponding state in the downstream matter for a given initial state. As the EoS is different, the upstream and downstream points lie on different curves. We plot the curves for the HM and its corresponding QM in the $X-p$ (chemical potential per particle versus pressure) plane (fig \ref{f4}). The initial point lies in the green HM curve, and for a given initial state, and solving the CA equation, we obtain a point lying in the burnt QM (red curve).
The range of the downstream curve for the GR case is significantly larger than the SR case, which signifies that more final states are available for the GR shock. Also, initially, the slope of the RL is larger for the SR shocks. The slope of the RL gives information about the change in pressure and density across the combustion front.

\begin{figure}
\centering
\hspace{0.5cm}
  \includegraphics[scale=0.8]{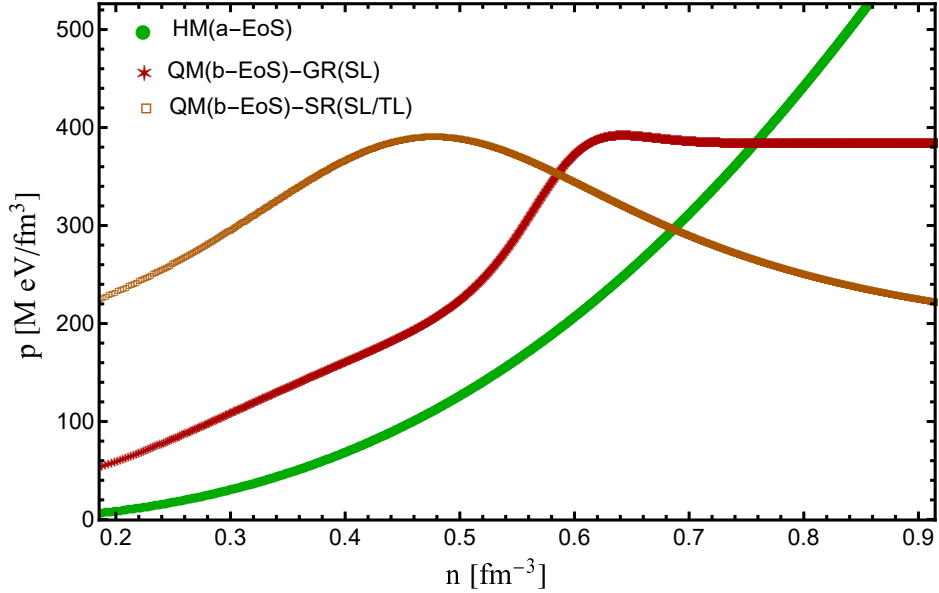}
  \caption{Pressure (p) as a function of baryon density (n) for HM (a-EoS) and their corresponding downstream QM (b-EoS) curves are shown in the figure. The burnt matter pressure first rises and then decreases, cutting their corresponding HM pressure curve at a particular n. }\label{f5}
\end{figure}

It is also helpful to plot how the pressure of the burnt matter varies with number density as we continuously change the initial state (HM). The green curve represents the HM curve in fig \ref{f5}. As we change the initial state, the final state also changes smoothly, as shown by the red curve (QM). We find that as the density and pressure of the initial state increase, first the density and pressure of the final state also increase for GRSL shocks; however, after a density of $0.6$ fm$^{-3}$ there is a small maximum after which the curve flattens out, and the pressure remains almost constant even if the density increases. For comparison, we have also plotted curves for SR shocks in the same figure.
For SR shock, the curve attains its maximum early around $0.5$ fm$^{-3}$, and then the pressure starts decreasing as density increases. Maxima of both the curves occurs at the same pressure but at different density values.

\begin{figure}
\centering
\hspace{0.5cm}
  \includegraphics[scale=0.8]{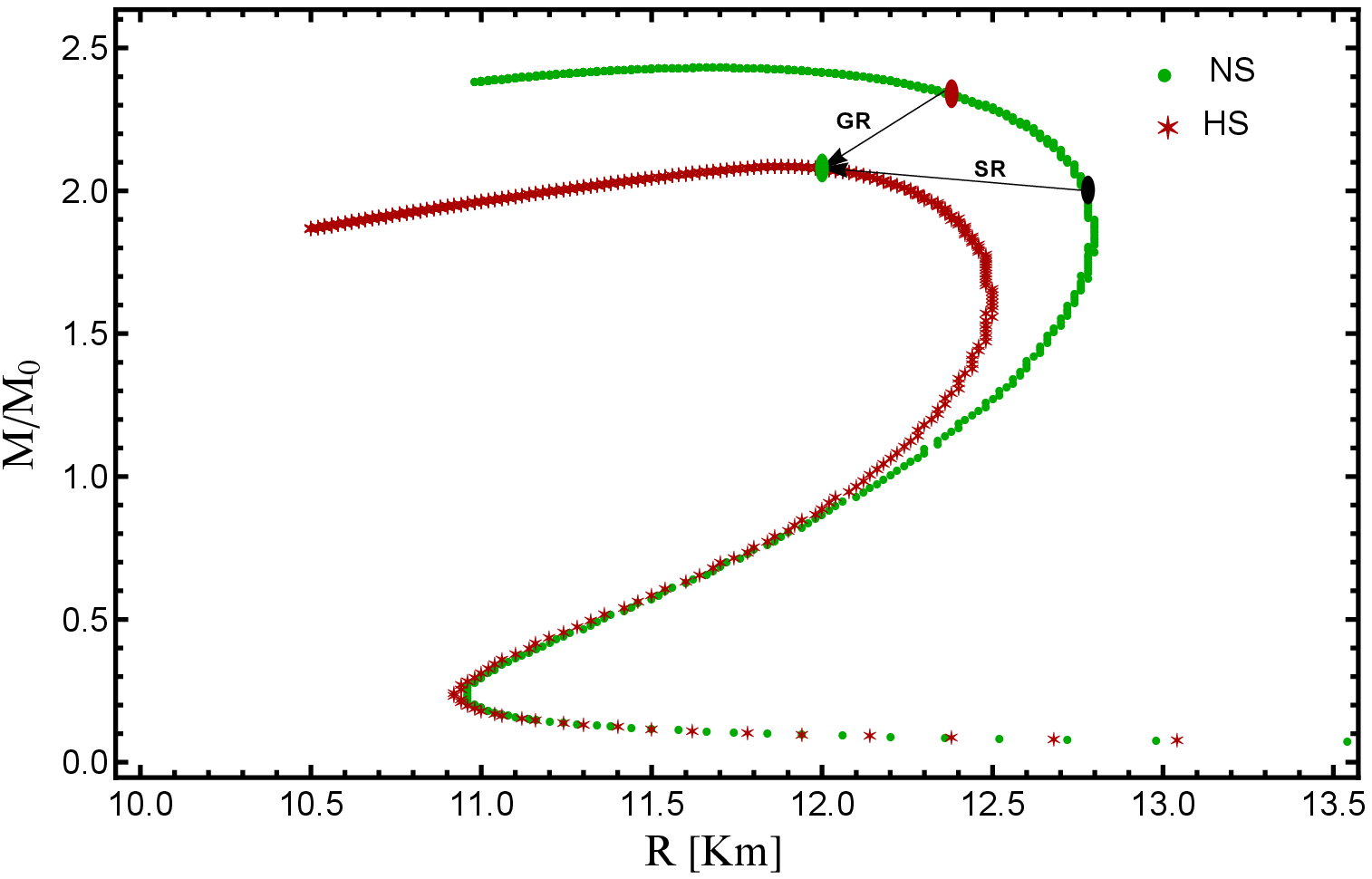}
  \caption{ Mass-Radius relations for NS and HS are plotted in the figure. Arrow indicates the maximum mass of phase transitioned HS combusted from an NS for GR and SR SL shocks.} \label{f6}
\end{figure}
 
 Once we know the pressure variation with density in the downstream matter, we can now solve the TOV equation along with the shock condition to generate stars that have undergone PT. We assume that PT happens only when the central density of the star is greater than the critical density for PT to occur. Therefore, for small stars where the central density is less than the critical density, PT does not take place, and we only have NSs. For relatively heavier stars where the central density is greater than the critical density, we have stars where combustion stars at the centre and happens till the point where the density is greater than the critical density. Therefore, we have an HS whose core is QM and is surrounded by a hadronic outer surface.
 
 The mass-radius diagram gives the sequences of the star's masses with their corresponding radius and is shown in fig \ref{f6}. It also gives the maximum mass that can be attained for a given EoS. The solid green curve gives the mass-radius sequence of the NS (governed by HM EoS). The maximum mass it can reach is about 2.42 $M_\odot$ corresponding to a radius of about 11.7 km. The mass-radius sequence of the HS is shown with a red curve which has a maximum mass of 2.08 $M_\odot$ corresponding to a radius of about 11.8 km. In this calculation, we have assumed that the HS is obtained from the parent NS due to PT of the core (where HM is converted to QM following the GR shock conditions). If we plot the mass-radius sequence of an HS obtained by solving the TOV equation for QM EoS (without shock-induced PT scenario), the maximum mass for the given EoS is about 2.08 $M_\odot$. For the particular choice of EoS, the calculations from GRSL CA gives the maximum mass of phase transitioned HS of about 2.08 $M_\odot$, which agrees well with the maximum mass obtained from plotting the HS EoS (without solving the CA equation). It is clear from the figure that the initial mass of NS is smaller than the daughter star mass for the SR case, which implies that for combustion to happen, we need an external source of energy. However, for the GR calculation, the initial star mass is slightly greater than the final daughter star mass, which is more likely to happen.  
 
\begin{figure}
\centering
\hspace{0.5cm}
  \includegraphics[scale=0.8]{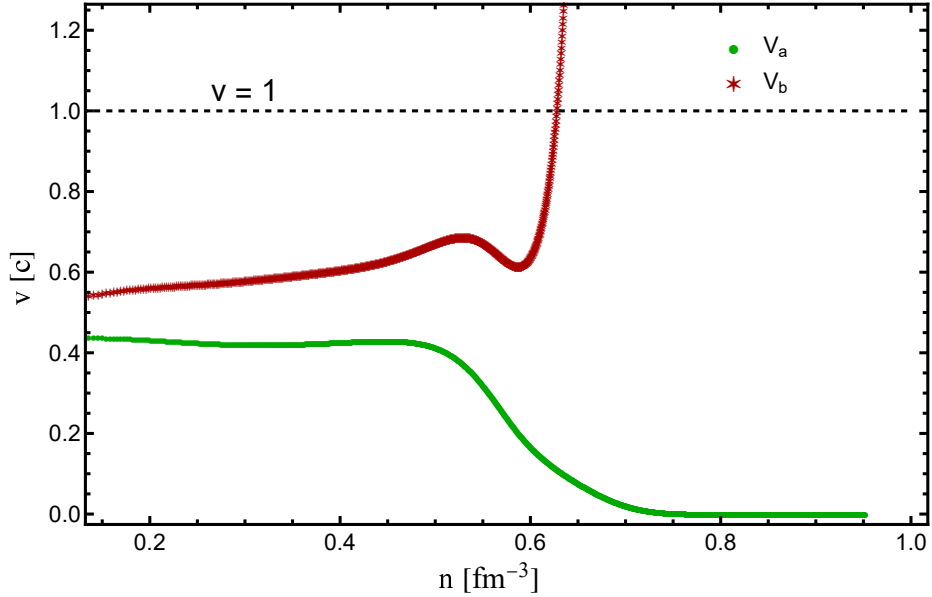}
  \caption{The upstream ($v_{a}$) for HM EoS (green-circle)  and downstream ($v_{b}$) for QM EoS (red-star) velocities are shown as a function of number density (n) for GRSL shocks. $v_{b}$ is seen to be greater than $v_{a}$ implying a detonation process. However, both velocities are subluminal in a low-density regime, but as density increases, the downstream matter velocity becomes superluminal, whereas the upstream velocity tends to zero, indicating a very fast combustion process.}
  \label{f3}
\end{figure}

The matter velocities across the front are a valuable tool for understanding the properties of shock-induced combustion. Combustion can be either detonation (fast burning where the combustion and the shock front almost coincide) or deflagration (slow burning). If the velocity of the burned matter is larger than unburnt matter, then PT corresponds to the detonation, whereas if its speed is lower than unburnt matter, then PT resembles the deflagration or slow combustion. This is expressed as
\begin{align}
v_{upstream}>v_{downstream} \Rightarrow Deflagration \nonumber \\
v_{upstream}<v_{downstream} \Rightarrow Detonation   \nonumber
\end{align}
In fig \ref{f3} we have shown how the upstream and downstream matter velocities vary with density during the combustion process. For low mass stars (whose central density is small), downstream matter velocity is larger than upstream velocity, indicating detonation. However, combustion is not instantaneous. However, for massive stars at the centre the combustion is almost instantaneous as the downstream matter velocity is very large. As it travels to low densities (outer region of the star), it attains a finite velocity. With GR calculation, the conversion process always remains a detonation, as seen from the figure.

In astrophysical scenarios, the matter velocities are usually SR. In the above discussion, we have only considered SL shocks, and even with it, we have encountered some density range where the downstream matter velocity becomes superluminous. Sometimes superluminous velocity arises only as a mathematical solution; however, it can even be interpreted as an instantaneous process. There can even be situations where the matter velocities become imaginary and non-physical \cite{taub,joel}. Starting with a given initial state and solving for the final state, one can categorize different physical and non-physical regions in an $\epsilon-p$ diagram, depending on the matter velocities. From eqn \ref{v1} and \ref{v2}, the condition under which matter velocities reach the velocity of light ($v=c=1)$ for the GRSL shocks can be determined. \\

a) condition for $v_a \approx 1$ is
\begin{align*}
    p_b &= \frac{h_1 + h_2 + h_3 + h_4}{A_2^4(A_1^2 + B_1^2)(A_1^2 (p_a + \epsilon_b) + B_1^2(\epsilon_b-\epsilon_a))}.
\end{align*}
b) and the condition for $v_b \approx 1$ is
 \begin{align*}
p_b = \frac{ k((A_1^4 B_1 B_2 w_a + A_2^4 B_1 B_2 \epsilon_b) - (p_a + \epsilon_b)A_1^3 B_2^2 A_2)}{A_2^3 B_1 (A_1 B_1 - k A_2 B_2)} \nonumber\\
 - \frac{k A_1 B_1^2 A_2^3 \epsilon_a}{A_2^3 B_1 (A_1 B_1 - k A_2 B_2)}
\end{align*} 
where,
\begin{align*}
    h_1 & = A_1^6 B_1^2 w_a^2\\
    h_2 & = A_2^4 B_1^4 \epsilon_a(\epsilon_a - \epsilon_b) \\
    h_3 & = A_1^4 A_2^4 p_a (p_a + \epsilon_b)\\
    h_4 & =A_1^2 A_2^4 B_1^2 (p_a(\epsilon_b-2\epsilon_a) - \epsilon_a \epsilon_b)\\
    k &= 1.00001
\end{align*}

and the condition for which the matter velocities become imaginary  for GR SL shocks are (same for both $v_a$ and $v_b$) 

\begin{align*}
\epsilon_a &<\epsilon_b  \textrm{ \&  } w_b^{2}A_2^{8} > A_1^{4}(2a_{11} + w_a^{2}A_1^{4}) \\
or \\
\epsilon_a &>\epsilon_b  \textrm{ \&  } w_b^{2}A_2^{8} < A_1^{4}(2a_{11} + w_a^{2}A_1^{4}) \\
\end{align*}

Similarly, for GR TL shocks the luminous velocity condition is given as \\

a) the condition for $v_a \approx 1$ becomes
    \begin{align*}
 p_b &= \frac{{A_2^{2}}({A_2^{2}}G1+{A_1^{2}}G3)}{{A_1^{2}} \left( {A_1}^6 {\epsilon_b}+{A_1}^4 G4 + {A_1^{2}}{B_1^{2}} G5 +{A_2^{2}}{B_1}^4 {p_a}\right)}
\end{align*}
b) and the condition for $v_b \approx 1$ becomes
\begin{align*}
p_b &= \frac{h {A_2^{2}} [{A_1^{2}} ({B_1^{2}}{\epsilon_b}+{B_2^{2}} {p_a}-{B_2^{2}}{\epsilon_a})-{A_2^{2}}{B_1^{2}}{p_a}]}{{A_1^{2}} ({A_1^{2}}{B_2^{2}}-{A_2^{2}}{B_1^{2}})} 
\end{align*}
where,
    \begin{align*}
    G1 &= ({A_1}^4{\epsilon_a}^2+2{A_1^{2}}{B_1^{2}}{p_a}{\epsilon_a}+{B_1}^4 {p_a}^2) \\
    G2 &=  (-{B_1^{2}}{p_a} {\epsilon_b}+{B_1^{2}} {\epsilon_a} {\epsilon_b}-{B_2^{2}}{p_a}^2 - 2 {B_2^{2}} {p_a}{\epsilon_a} -{B_2^{2}}{\epsilon_a}^2)\\
    G3 &= -{A_1}^4 {\epsilon_a} {\epsilon_b}+{A_1^{2}}G2 + {B_1}^4{p_a}{\epsilon_b} \\
    G4 &= -({A_2^{2}}{\epsilon_a}-2 {B_1^{2}} {\epsilon_b})\\
    G5 &=  (-{A_2^{2}}{p_a}+{A_2^{2}}{\epsilon_a}+{B_1^{2}}{\epsilon_b})\\
    h & = Constant.
    \end{align*}
and imaginary velocity condition is
\begin{align*}
A_2 p_a &< A_1 p_b  \textrm{  \&  } a_{21} > b_{21} \\
or \\
A_2 p_a &> A_1 p_b  \textrm{   \&  } a_{21} < b_{21}  .
\end{align*}

\begin{figure}
\centering
\hspace{0.5cm}
  \includegraphics[scale=0.8]{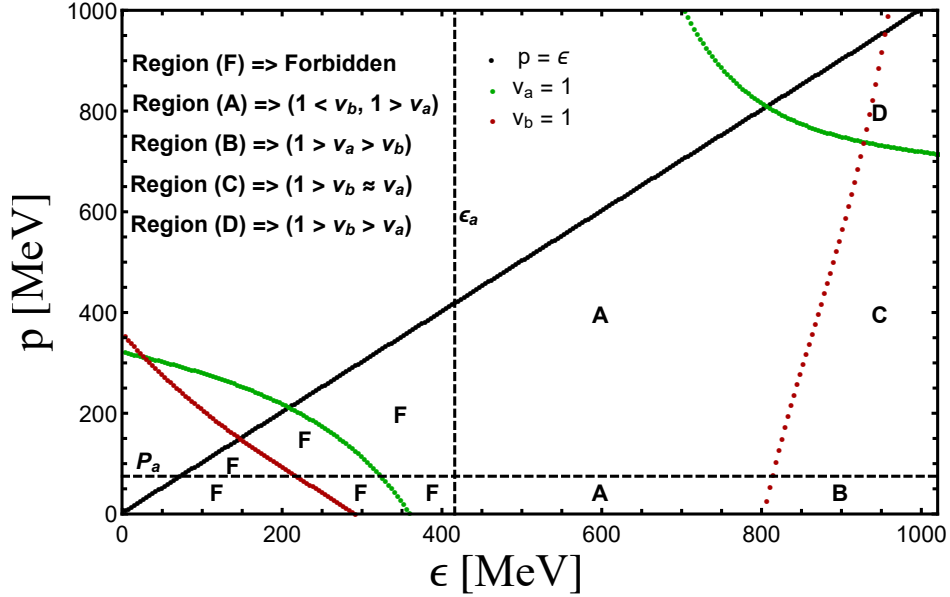}
  \caption{Final allowed states p versus $\epsilon$ diagrams are shown for shock-induced phase transition for a particular initial state ($\epsilon_a$,$p_a$) for GRSL shocks. The black line shows the causality line. The dark red and dark green curve shows the luminal condition for downstream and upstream velocity, respectively. Region A indicates the superluminal velocity of downstream matter favoring the detonation process, while in Region B, C, and D, both upstream and downstream velocities are subluminal, whereas Region F is forbidden in which both upstream and downstream matter velocities becomes imaginary. Region C and D also favor the detonation process, but Region B supports the deflagration process.}
   \label{region}
\end{figure}

In fig \ref{region} we have plotted the range for which matter velocities are subluminous, superluminous, and completely imaginary for the SL GR shocks. The solid black line indicates the boundary of causality \cite{rhoades,gorenstein}. The region below it is causally connected. The solid green and red line marks the boundary for $v_a$ and $v_b$ to luminal simultaneously in the GR SL case. Once we know the boundary, we can mark the region of subluminal (Region B, C, and D), superluminal (region A), and forbidden (complex) velocities. We have drawn a similar figure for TL GR shocks where the different marked regions are shown (Appendix).

\section{Summary and Conclusion}

In this paper, we have studied the different properties of GR shock jump conditions. We started with the given hydrodynamic equations for fluid, and from there, we first derived the GR jump conditions for SL and TL shocks. From the jump conditions, we have further derived the TA/CA conditions. We found that the matter velocities across the discontinuity for SL and TL shocks are not inversely related (as was the case for SR shocks).

Once the jump conditions and CA are derived, we then analyze the situation when the discontinuity is not very large. For the SR case, we found that it refers to the condition of the isentropic process. We did a similar calculation for GR shocks; however, to analyze weak shocks in GR, we also required information about the metric potentials. To obtain the metric potential, we choose our system to be an NS. With the condition for small discontinuity, we found that it refers to an isentropic process even for GR shocks in an NS. We then discussed general shock waves and checked whether, at CJ point, the velocity of downstream matter becomes equal to the speed of sound locally. For the SR SL shock, the condition holds; however, for SR TL shocks, it fails but is just ($\sqrt{1-c_s^2}$). For the NS system, we checked whether anywhere in the star for GR shocks at CJ point; the downstream matter becomes equal to the speed of sound locally. However, such a point is not found for the GR shocks. The downstream matter 4-velocity is always less than 4-velocity of sound in the star.     

Next, we checked how GR calculation differs from the SR calculation if we are to treat the PT happening from HM to QM as shock-induced combustion. We have discussed results only for the SL shocks as they are more common in astrophysical scenarios.
We found that combustion adiabat has a much wider range for GR shocks than for SR shocks. However, the maximum pressure attained by both remains the same. This means that the maximum mass of the phase transformed daughter QS for both GR and SR calculation remains the same. However, for SR shocks for the combustion to happen, one needs some external source as the process is endothermic. For the GR shock, the combustion process is exothermic and therefore is much more likely. We found that for GR shocks, the combustion process is always a detonation, and at high density, it is almost instantaneous. We also found that for GR shocks for a given initial state, there can be regions in the star where the matter velocities can be superluminous, and there are even regions where such combustion is forbidden.

Therefore, to treat astrophysical shock problems kinematically, one needs to have GR jump condition and CA equation. GR calculation provides a much more robust and physical picture. There are many future astrophysical problems that can be addressed with our formalism of GR shocks. A possible extension would be to study oblique shock, and then from there, one can calculate the particle acceleration due to shocks. Most of the particle acceleration calculation involves relativistic and non-relativistic formalism; however, as clear from our analysis, more robust results can be expected from GR analysis.

\section{Acknowledgments}
The authors thank the Indian Institute of Science Education and Research Bhopal for providing all the research and infrastructure facilities. We would also like to thank Aditya Sharma and Shailendra Singh for the fruitful discussions. 

\section*{Data Availability}
This paper has no additional data as it is a theoretical work.

\appendix
\section{}
Plots showing the results of GR TL shocks under same initial conditions for GR SL shocks discussed in the main text. The plots are done with same HM and QM EoSs as done in the main text.

\begin{figure}[h]
\centering
\hspace{0.5cm}
  \includegraphics[scale=0.7]{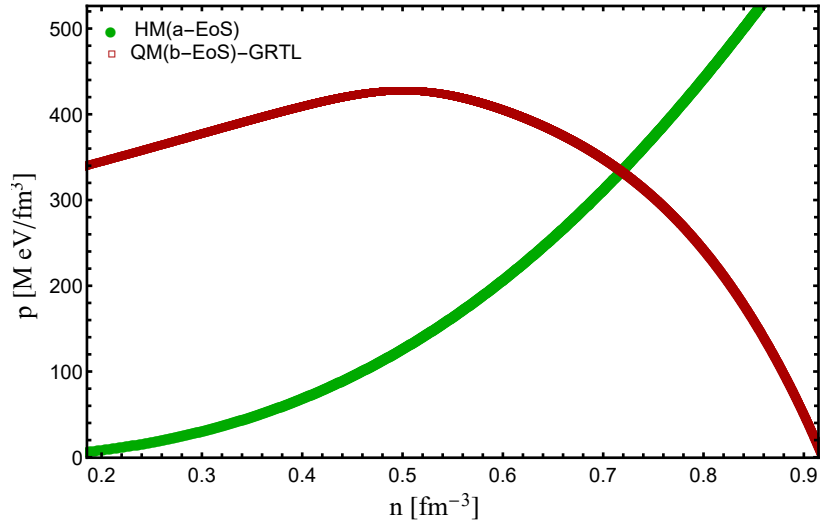}
  \caption{Pressure (p) as a function of baryon density (n) for HM (a-EoS) and their corresponding downstream QM (b-EoS) curves are shown in the figure for GR TL shock. The burnt matter pressure first rises and then decreases, giving us a maximum QM pressure.}
  \label{grtln}
\end{figure}

\begin{figure}[h]
\centering
\hspace{0.5cm}
  \includegraphics[scale=0.7]{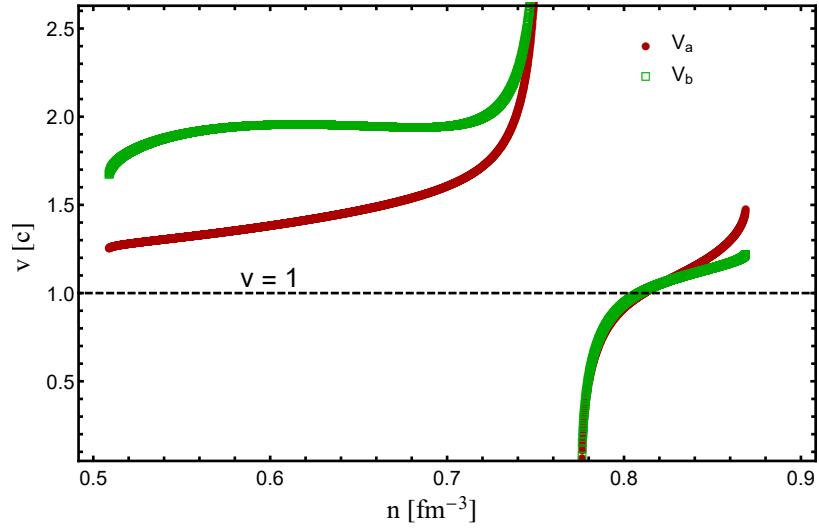}
  \caption{The upstream ($v_{a}$) for HM EoS (green-circle)  and downstream ($v_{b}$) for QM EoS (red-star) velocities are shown as a function of number density (n) for GR TL shocks. $v_{b}$ is seen to be greater than $v_{a}$ implying a detonation process. However, both velocities are superluminal in a low-density regime, but as density increases and crosses a particular density value, the downstream and upstream matter velocity becomes equal as well as subluminal for a small density range.}
  \end{figure}

\begin{figure}
\centering
\hspace{0.5cm}
  \includegraphics[scale=0.7]{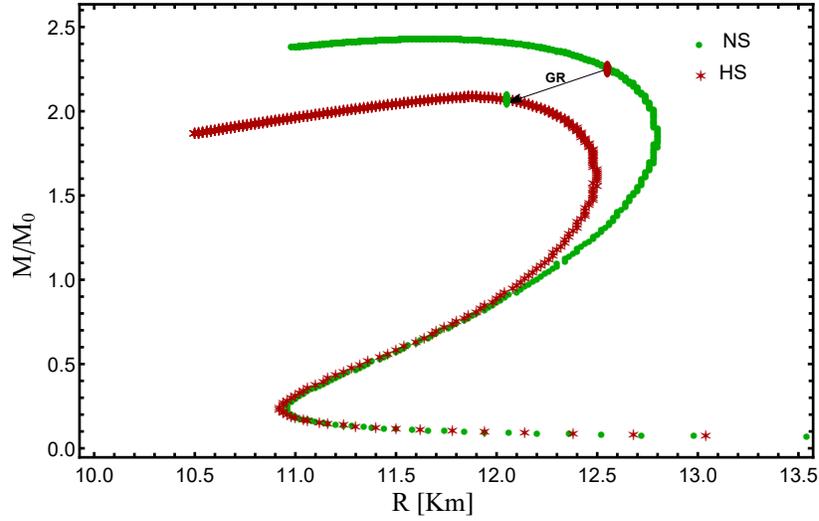}
  \caption{ Mass-Radius relations for NS and HS are plotted in the figure. Arrow indicates the maximum mass of phase transitioned HS combusted from an NS for GR TL shock corresponding to the maximum quark pressure obtained from the fig \ref{grtln}.}
  \end{figure}

  \begin{figure}
\centering
\hspace{0.5cm}
  \includegraphics[scale=0.7]{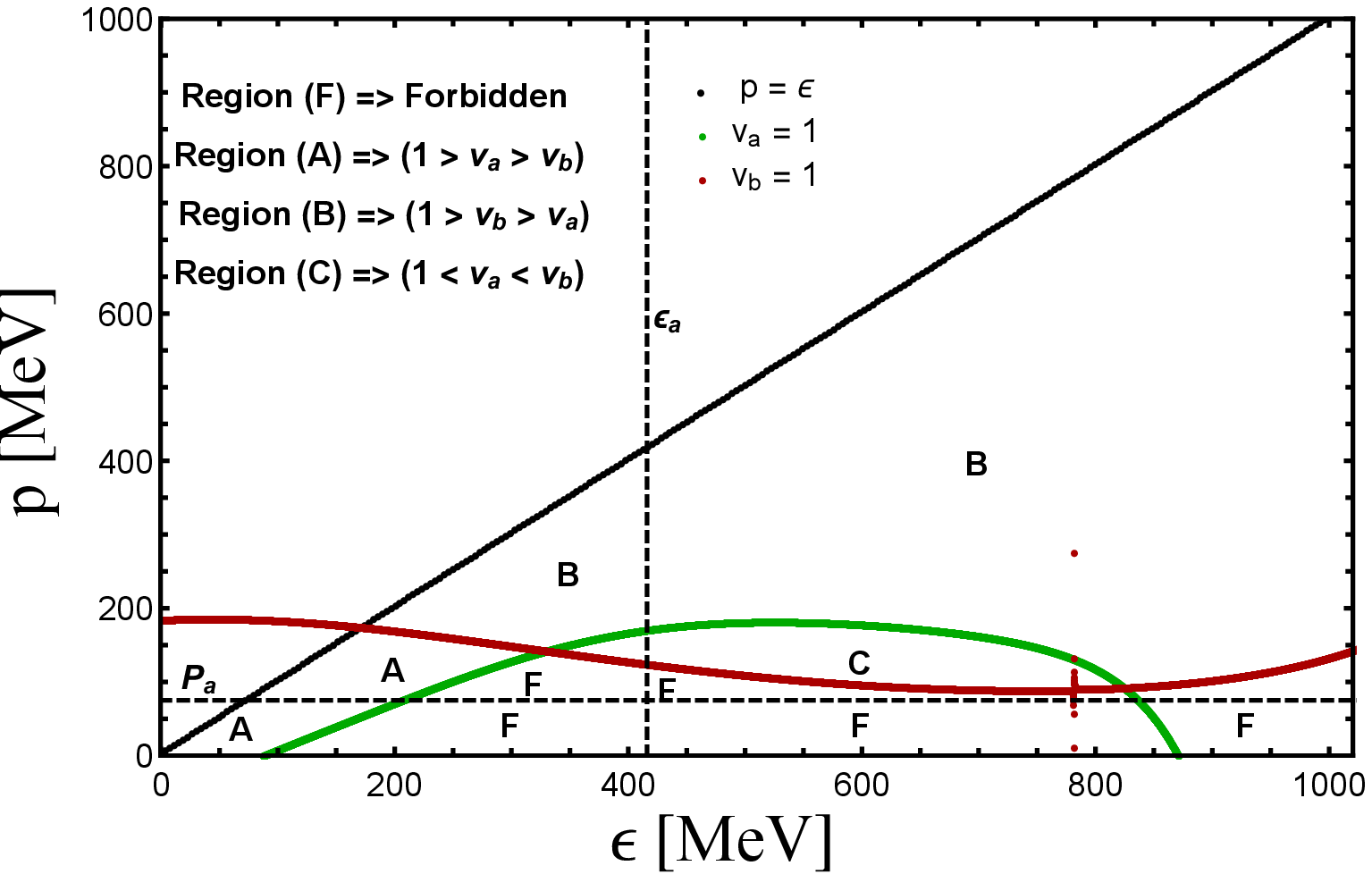}
  \caption{Final allowed states in the p versus $\epsilon$ diagrams are shown for shock-induced phase transition for a particular initial state ($\epsilon_a$,$p_a$) for GR TL shocks. The black line shows the causality line. The dark red and dark green curve shows the luminal condition for downstream and upstream velocity, respectively. Region A and B indicate the subluminal velocity of downstream matter favoring detonation and deflagration process, respectively. In Region C, both upstream and downstream velocities are superluminal, whereas Region F is forbidden in which both upstream and downstream matter velocities becomes imaginary.}
\end{figure}

\end{document}